# Collapse to Black Holes in Brans-Dicke Theory: I. Horizon Boundary Conditions for Dynamical Spacetimes

Mark A. Scheel

*Center for Radiophysics and Space Research*
*and Department of Physics,*
*Cornell University, Ithaca, New York 14853*

Stuart L. Shapiro and Saul A. Teukolsky

*Center for Radiophysics and Space Research*
*and Departments of Astronomy and Physics,*
*Cornell University, Ithaca, New York 14853*

ABSTRACT: We present a new numerical code that evolves a spherically symmetric configuration of collisionless matter in the Brans-Dicke theory of gravitation. In this theory the spacetime is dynamical even in spherical symmetry, where it can contain gravitational radiation. Our code is capable of accurately tracking collapse to a black hole in a dynamical spacetime arbitrarily far into the future, without encountering either coordinate pathologies or spacetime singularities. This is accomplished by truncating the spacetime at a spherical surface inside the apparent horizon, and subsequently solving the evolution and constraint equations only in the exterior region. We use our code to address a number of long-standing theoretical questions about collapse to black holes in Brans-Dicke theory.

# I. INTRODUCTION

In recent years, there has been renewed interest in scalar-tensor theories of gravitation. One reason is that these theories are important for cosmological inflation models[1], in which the scalar field allows the inflationary epoch to end via bubble nucleation without the need for fine-tuning cosmological parameters (the "graceful exit" problem). In addition, scalar-tensor gravitation ("dilaton gravity") arises naturally from the low-energy limit of superstring theories[2, 3]. Finally, with the construction of LIGO, it may be possible to test scalar-tensor theories to high precision[4] by looking for monopole and dipole gravitational radiation from astrophysical sources.

Quite apart from their potential physical significance, scalar-tensor theories play another very useful role: they provide an ideal laboratory for testing new algorithms for numerical relativity. In general relativity, numerical methods for treating spacetimes containing gravitational radiation require at least two spatial dimensions, since a time-varying quadrupole moment is needed to produce gravitational waves. In scalar-tensor theories, one can study many of the same strong-field phenomena that occur in general relativity, including gravitational radiation and dynamical black holes, while still working in spherical symmetry.

We have developed a numerical code that solves the coupled matter and gravitational field equations for the evolution of a spherically symmetric configuration of noninteracting particles in Brans-Dicke[5] gravitation, the simplest of the scalar-tensor theories. We use this code to study gravitational collapse to a black hole in Brans-Dicke theory. This process has been discussed extensively in the literature[6], but these studies have been limited to addressing the final state of the black hole after collapse, or have used linearized approximations of the field equations. Other than an early simulation by Matsuda and Nariai[7], it is only very recently[4] that this process has been calculated in any detail.

In constructing numerical models of gravitational collapse in Brans-Dicke theory, we have been forced to address the same difficulty that has plagued the field of numerical relativity for the last 30 years: how does one handle the spacetime singularity at the origin that inevitably develops during the formation of a black hole?

The traditional approach has been to utilize the "many-fingered time" gauge freedom of general relativity to avoid the singularity altogether. Specifically, one chooses coordinates such that the passage of proper time grinds to a halt near the origin before the singularity appears, while weak-field regions of spacetime farther from the origin evolve farther into the future. This singularity-avoiding (SA) method works well for short times, but eventually pathologies develop in the transition region between the "frozen" interior and the "evolving" exterior. These typically take the form of steep gradients or spikes in the metric functions, and will eventually cause the numerical code to crash[8]. Countermeasures such as increasing the grid resolution produce little improvement because the pathologies increase exponentially with time.

Our solution to this problem is to use an apparent horizon boundary condition (AHBC) method after the formation of a black hole. The basic idea of this approach is to truncate a black hole spacetime at a surface inside the apparent horizon (AH) that cannot causally influence the exterior. One then discards the singular interior entirely, and only evolves the physically relevant exterior.

Seidel and Suen[9] have implemented this idea in general relativity for the case of a black hole with a Klein-Gordon field in spherical symmetry. Their method involves a coordinate system that is locked to the AH, so that the coordinate speed of radially outgoing light rays inside the AH is negative. This enables them to use a causal difference scheme, similar to the Causal Reconnection Scheme of Alcubierre and Schutz[10], to solve evolution equations in such a way that information does not escape from the black hole, and no explicit boundary condition is needed on the AH.

Our AHBC method is different from that of Seidel and Suen. Although we also use a coordinate system that is locked to the AH, we solve the wave equation for the Brans-Dicke scalar field using an implicit differencing scheme motivated by the work of Alcubierre[11]. In addition, we solve for the metric variables using the elliptic constraint equations rather than the evolution equations. We obtain the required boundary conditions for this approach from asymptotic flatness, properties of the apparent horizon, and by solving a single evolution equation only on the AH, as explained in Section IV.

With our code we are able to follow accurately Brans-Dicke collapse to black holes and the associated generation of monopole gravitational waves. We are able to integrate the equations to arbitrarily late times into the future, when all of the radiation has propagated out to large distances and the central black hole has settled down to final equilibrium. Using our code we are able to resolve a number of long-standing, theoretical questions about collapse in Brans-Dicke theory. We are also able to refine a promising technique for evolving black hole spacetimes with radiation by integrating only in the observable regions of spacetime.

This paper is primarily concerned with numerical methods. The reader not interested in the numerical details should read section II and then skip to Paper II[12], in which we discuss Brans-Dicke gravitation in more detail, and we show how black holes formed in Brans-Dicke theory behave differently than those in general relativity.

## II. BASIC EQUATIONS

### A. Brans-Dicke Theory

The action for Brans-Dicke gravitation is[5]

$$I = \int \mathcal{L}_{\rm BD}(-g)^{1/2} d^4x, \tag{2.1}$$

where the Lagrangian density is

$$\mathcal{L}_{\rm BD} = g^{ab} R_{ab}\phi + \frac{16\pi}{c^4}\mathcal{L} - \frac{\omega}{\phi} g^{ab}\partial_a\phi\partial_b\phi. \tag{2.2}$$

The coupling constant $\omega$ is dimensionless, and the scalar field $\phi$ has dimensions of $G^{-1}$, where $G$ is Newton's gravitational constant. The Lagrangian density $\mathcal{L}$ for matter and nongravitational fields depends on the metric $g^{ab}$ but not on $\phi$. The Ricci tensor $R_{ab}$ is related to the metric in the usual way. In general relativity, the third term in Eq. (2.2) is absent, and one sets $\phi = G^{-1}$. In other scalar-tensor theories[13], the Lagrangian is more complicated. In particular, the coupling parameter $\omega$ can be a function of $\phi$.

In general relativity we are free to choose our units of mass and time such that $G = c = 1$. In Brans-Dicke theory, the inverse of the scalar field $\phi^{-1}$ plays the role of $G$, so multiplying $\phi$ by a global scaling factor changes the unit of mass. We choose units such that

$$c = \phi_\infty = 1, \tag{2.3}$$

where $\phi_\infty$ is the value of the scalar field far from any sources.

Variation of Eq. (2.1) with respect to $g^{ab}$ and $\phi$ yields the Brans-Dicke field equations, which can be written in the form

$$\Box\phi = \frac{8\pi T}{3+2\omega}, \tag{2.4}$$

$$G_{ab} = 8\pi T'_{ab}, \tag{2.5}$$

where

$$8\pi T'_{ab} \equiv \frac{1}{\phi}\left[8\pi T_{ab} + \frac{\omega}{\phi}\left(\nabla_a\phi\nabla_b\phi - \tfrac{1}{2}g_{ab}\nabla_a\phi\nabla^a\phi\right) + \nabla_a\nabla_b\phi - g_{ab}\Box\phi\right]. \tag{2.6}$$

Here $\nabla$ denotes covariant differentiation with respect to the metric $g_{ab}$, $\Box$ is the covariant Laplacian $\nabla^a\nabla_a$, and $G_{ab}$ is the usual Einstein tensor. The symmetric tensor $T_{ab}$ is the energy-momentum tensor for matter and nongravitational fields, and $T$ is its trace:

$$T_{ab} \equiv g_{ab}\mathcal{L} - 2\frac{\delta\mathcal{L}}{\delta g^{ab}}, \tag{2.7}$$

$$T \equiv T^a{}_a = T_{ab}g^{ab}, \tag{2.8}$$

$$\nabla_b T^{ab} = 0. \tag{2.9}$$

Although we have written the field equations (2.5) in a form that resembles Einstein's equations, we emphasize that Brans-Dicke gravitation is *not* the same as general relativity with a Klein-Gordon scalar field. The difference is the factor of $\phi$ in the first term in the Lagrangian density (2.2), which leads to second derivatives of $\phi$ in the field equations (2.5). Physically, this manifests itself as a violation of the weak equivalence principle for massive bodies[14]. This is discussed in more detail in Paper II.

Notice that the matter stress-energy is conserved (Eq. 2.9), even though $T^{ab}$ is not equal to $8\pi G^{ab}$ (the quantity $\nabla_b T'^{ab}$ also vanishes because of the Bianchi identity $\nabla_b G^{ab} = 0$). As a result, the equations of motion of matter do not involve the scalar field; test particles move on geodesics of the metric. Notice also that it is the trace of the *matter* stress-energy tensor $T_{ab}$, not the effective tensor $T'_{ab}$, that appears in the wave equation (2.4).

In vacuum, $\phi =$ constant is a solution of Eq. (2.4) for any $\omega$. In this case, $T'_{ab}$ reduces to $T_{ab}$ and the field equations (2.5) become Einstein's equations. Therefore, any vacuum solution of Einstein's equations is also a vacuum solution of the Brans-Dicke equations. In addition, many (but not all) Brans-Dicke solutions with $|\omega| \to \infty$ have $\phi \to$ constant and obey Einstein's equations. It is therefore said (but not rigorously correct—See Paper II) that Brans-Dicke theory reduces to general relativity in the limit $|\omega| \to \infty$.

## B. Equations (2.5) in (3+1) form

Adopting the usual ADM[15] $(3+1)$ decomposition, we introduce a set of spacelike hypersurfaces, or time slices, and a timelike vector field $n^a$ normal to these hypersurfaces. We adopt the convention that the 4-metric $g_{ab}$ has signature $(-+++)$, so that

$$n^a n_a = -1. \tag{2.10}$$

Spatial distances on a particular time slice are measured by the 3-metric $\gamma_{ab}$, defined by

$$\gamma_{ab} \equiv g_{ab} + n_a n_b. \tag{2.11}$$

The extrinsic curvature $K_{ab}$ describes the rate of change of the three-metric along $n^a$ (the "time" direction):

$$K_{ab} \equiv -\tfrac{1}{2}\pounds_n \gamma_{ab}. \tag{2.12}$$

Here $\pounds$ denotes a Lie derivative. The field equations are split into spatial constraints that relate $K_{ab}$ and $\gamma_{ab}$ on each time slice, and first-order (in time) evolution equations that take $K_{ab}$ and $\gamma_{ab}$ from one slice to the next.

We work in spherical symmetry, and we choose the maximal time slicing and isotropic radial coordinate conditions. This gauge is defined by the isotropic line element

$$ds^2 = -(\alpha^2 - A^2\beta^2)\,dt^2 + 2A^2\beta\,dr\,dt + A^2(dr^2 + r^2\,d\Omega^2) \tag{2.13}$$

and the maximal slicing condition

$$K = K_{,t} = 0. \tag{2.14}$$

Here $K$ is the trace of the extrinsic curvature $K_{ab}$. The 3-dimensional metric $\gamma_{ij}$ on each $t$ = constant time slice is given by

$$ {}^{(3)}ds^2 = A^2(dr^2 + r^2\, d\Omega^2). \tag{2.15}$$

Eqs. (2.10) and (2.11) imply

$$ n_a = (-\alpha, 0, 0, 0). \tag{2.16}$$

The lapse function $\alpha$ measures the ratio of proper time to coordinate time for a normal observer:

$$ d\tau = \alpha\, dt. \tag{2.17}$$

The shift $\beta$ is the velocity of the spatial coordinates with respect to normal observers. It is a vector quantity, but in spherical symmetry only the radial component is nonzero ($\beta^a = 0$ for $a \neq r$), so we write $\beta \equiv \beta^r$. The shift is crucial for our AHBC method: in order for a coordinate grid point to have no causal effect on the region exterior to that point, the coordinate speed of radially outgoing photons must not be positive, and this requires a (positive) shift.

The maximal slicing condition (2.14) is important for our conventional SA method because it causes the lapse function to become small in the strong-field region of a spatial slice that is about to hit a singularity. This slows down the passage of proper time in this region while other regions on the slice propagate farther into the future. In a typical black hole spacetime, maximal slices will never encounter the singularity at $r = 0$. For the AHBC method, maximal slicing is not at all necessary, but it is convenient because it eliminates a component of $K_{ab}$ from the equations.

We choose the isotropic spatial gauge primarily for convenience. The most general spherically symmetric metric can be written in the form

$$ ds^2 = -(\alpha^2 - A^2\beta^2)\, dt^2 + 2A^2\beta\, dr\, dt + A^2 dr^2 + B^2 r^2\, d\Omega^2. \tag{2.18}$$

The isotropic gauge condition $A = B$ eliminates $B$ from the field equations. This choice is not necessary for either the SA or AHBC methods. However, there are gauge choices (for instance, $\beta \equiv 0$) that would spoil the AHBC method.

Note that Seidel and Suen[9] do not use isotropic coordinates, but instead choose the metric (2.18) and the gauge condition $\partial A/\partial t = 0$. This choice is useful because the proper distance between two coordinate radii remains fixed in time, but it does not seem to be necessary for the success of the AHBC method, at least in spherical symmetry. In fact, it tends to complicate the analysis after matching onto the SA method, especially in the linearized regime where it produces a frozen coordinate wave that adds gauge terms to the metric variables.

In maximal isotropic gauge, the field equations (2.5) break up into two evolution equations,

$$ \frac{A_{,t}}{A} - \beta\frac{A_{,r}}{A} = \frac{\beta}{r} - \frac{1}{2}\alpha K_{\mathrm{T}}, \tag{2.19}$$

$$ K_{\mathrm{T},t} - \beta K_{\mathrm{T},r} = \alpha\left[8\pi {S^r}_r{}' + \frac{3}{4}K_{\mathrm{T}}^2 - \frac{A_{,r}}{A^3}\left(\frac{A_{,r}}{A} + \frac{2}{r}\right)\right] - \frac{2\alpha_{,r}}{A^2}\frac{(Ar)_{,r}}{Ar}, \tag{2.20}$$

and four spatial constraint equations

$$ \frac{1}{r^2}\left(r^2(A^{1/2})_{,r}\right)_{,r} = -\frac{A^{5/2}}{4}\left(8\pi\rho' + \frac{3}{4}K_{\mathrm{T}}^2\right), \tag{2.21}$$

$$ K_{\mathrm{T},r} = 8\pi S_r' - 3K_{\mathrm{T}}\frac{(Ar)_{,r}}{Ar}, \tag{2.22}$$

$$ \frac{1}{\alpha A^3 r^2}\left(Ar^2\alpha_{,r}\right)_{,r} = \frac{3}{2}K_{\mathrm{T}}^2 + 8\pi\rho' + 4\pi T', \tag{2.23}$$

$$ r\left(\frac{\beta}{r}\right)_{,r} = -\frac{3}{2}\alpha K_{\mathrm{T}}. \tag{2.24}$$

Here and elsewhere in this paper, commas denote partial derivatives and

$$K_{\rm T} \equiv 2K^{\theta}{}_{\theta} = 2K^{\phi}{}_{\phi}. \tag{2.25}$$

Eqs. (2.21) and (2.22) are the Hamiltonian and momentum constraints. Eqs. (2.23) and (2.24) result from the maximal slicing condition $K_{\rm T} = -K^r{}_r$ and the isotropic coordinate condition $A = B$, respectively. It is important to note that Eqs. (2.19)–(2.24) are not all independent: one may use the Bianchi identities to eliminate two of the six equations.

The effective source terms appearing in Eqs. (2.19)–(2.24) are defined by

$$\begin{aligned} T' &\equiv g^{ab} T'_{ab}, \\ \rho' &\equiv n^a n^b T'_{ab}, \\ S'_a &\equiv \gamma^b_a n^c T'_{bc}, \\ S'_a &\equiv \gamma^b_a n^c T'_{bc}, \\ S'_{ab} &\equiv \gamma^c_a \gamma^d_b T'_{cd}, \end{aligned} \tag{2.26}$$

where $T'_{ab}$ is given by Eq. (2.6). Similarly, matter source terms $\rho$, $T$, $S_a$, and $S_{ab}$ (without the prime) are defined as in Eqs. (2.26), with the *unprimed* $T_{ab}$ appearing on the right hand side of each equation. If one replaces the primed source terms in Eqs. (2.19)–(2.24) with their unprimed counterparts, one recovers the Einstein equations.

### C. Scalar Field

To solve the scalar wave equation (2.4) numerically, it is useful to define the variables

$$\Pi \equiv -\pounds_n \phi, \tag{2.27}$$

$$\Phi \equiv \phi_{,r}. \tag{2.28}$$

In our coordinate system, Eq. (2.27) reduces to

$$\phi_{,t} = \beta\Phi - \alpha\Pi, \tag{2.29}$$

and the scalar wave equation (2.4) becomes

$$\Phi_{,t} = \beta\Phi_{,r} + \beta_{,r}\Phi - (\alpha\Pi)_{,r}, \tag{2.30}$$

$$\Pi_{,t} = \beta\Pi_{,r} + \frac{8\pi T\alpha}{3+2\omega} - \frac{1}{A^3 r^2}\left(Ar^2\alpha\Phi\right)_{,r}. \tag{2.31}$$

Using $\Pi$ rather than $\phi_{,t}$ as a dynamical variable eliminates explicit time derivatives of $\alpha$ and $\beta$ in Eq. (2.31). Furthermore, like the extrinsic curvature $K_{ab}$, the quantity $\Pi$ is defined geometrically (by Eq. 2.27) on each time slice, independent of how one gets from one slice to the next: in other words, its value on a given slice is independent of $\alpha$ and $\beta$. This is not true for $\phi_{,t}$. This is important when making the transition from the SA method to the AHBC method: the lapse and shift are discontinuous across the time slice that defines this transition.

The quantity $\Phi$ is a useful dynamical variable for the AHBC method because it eliminates explicit second derivatives from the wave equation. However, in the SA method it is sufficient to use $\phi$ as a dynamical variable and to explicitly compute $\phi_{,r}$ and $\phi_{,rr}$ by finite differencing when necessary.

Evaluating the effective source terms (2.26) using Eq. (2.6) and the metric (2.13) yields

$$8\pi T'\phi = \frac{16\pi T}{(2+3/\omega)} + \frac{\omega}{\phi}\left(\Pi^2 - \frac{\Phi^2}{A^2}\right), \tag{2.32a}$$

$$8\pi\rho'\phi = 8\pi\rho + \frac{\omega}{2\phi}\left(\Pi^2 + \frac{\Phi^2}{A^2}\right) + \frac{1}{A^3 r^2}\left(\Phi A r^2\right)_{,r}, \tag{2.32b}$$

$$8\pi {S^r}_r{}'\phi = 8\pi {S^r}_r - \frac{8\pi T}{(3+2\omega)} + \frac{\omega}{2\phi}\left(\Pi^2 + \frac{\Phi^2}{A^2}\right) + \Pi K_{\rm T} + \frac{1}{A}\left(\frac{\Phi}{A}\right)_{,r}, \tag{2.32c}$$

$$8\pi S_r'\phi = 8\pi S_r - \frac{\Phi\Pi\omega}{\phi} - \Pi_{,r} - \Phi K_{\rm T}. \tag{2.32d}$$

## D. Linearized Equations in Vacuum

In the weak-field regime, we can use linearized theory to describe the gravitational field. By matching our numerical variables to the linearized solution, we can determine the gravitational radiation seen by an observer at infinity, and we can set boundary conditions at the outer edge of our finite-difference grid. These boundary conditions, including the ones imposed on elliptic equations, are valid even while the wave is passing through the boundary. Such a matching technique was introduced in numerical relativity by Abrahams and Evans[16].

Define the new variables

$$a \equiv A - 1, \tag{2.33}$$

$$\epsilon \equiv \alpha - 1, \tag{2.34}$$

$$\xi \equiv \phi - 1. \tag{2.35}$$

If we set all matter terms to zero in Eqs. (2.19)–(2.24) and Eqs. (2.29)–(2.32), and if we keep only terms linear in $a$, $K_{\rm T}$, $\epsilon$, $\beta$, $\xi$, $\Phi$, and $\Pi$, we obtain

$$\xi_{,tt} = \frac{1}{r^2}\left(r^2\xi_{,r}\right)_{,r}, \tag{2.36}$$

$$\Pi = -\xi_{,t}, \tag{2.37}$$

$$\Phi = \xi_{,r}, \tag{2.38}$$

$$a_{,t} = \frac{\beta}{r} - \frac{1}{2}K_{\rm T}, \tag{2.39}$$

$$K_{{\rm T},t} = -\frac{2\epsilon_{,r}}{r} - \frac{2a_{,r}}{r} + \Phi_{,r}, \tag{2.40}$$

$$a_{,rr} = -\frac{2a_{,r}}{r} + \frac{\Phi_{,r}}{2} - \frac{\Phi}{r}, \tag{2.41}$$

$$K_{{\rm T},r} = -\frac{3K_{\rm T}}{r} - \Pi_{,r}, \tag{2.42}$$

$$\left(r^2\epsilon_{,r}\right)_{,r} = \left(r^2\xi_{,r}\right)_{,r}, \tag{2.43}$$

$$r\left(\frac{\beta}{r}\right)_{,r} = -\frac{3}{2}K_{\rm T}. \tag{2.44}$$

If only outgoing waves are present, the solution of Eq. (2.36) is

$$r\xi = f(t-r), \tag{2.45}$$

where $f$ is an arbitrary function of $(t-r)$. We now insert Eq. (2.45) into Eqs. (2.39)–(2.44), and impose the boundary conditions $\epsilon = a = 0$ at $r = \infty$. We find the solution:

$$A = 1 + a = 1 + \frac{M_{\rm T}}{r} - \frac{f(t-r)}{2r}, \tag{2.46}$$

$$\alpha = 1 + \epsilon = 1 + \frac{C(t)}{r} + \frac{f(t-r)}{r}, \tag{2.47}$$

$$\beta = \frac{rK_{\rm T}}{2} - \frac{f'(t-r)}{2}, \tag{2.48}$$

$$K_{\rm T} = \frac{f'(t-r)}{r} + \frac{3f(t-r)}{r^2} + \frac{1}{r^3}\int_0^t \left(2M_{\rm T} + 3f(\tilde{t}-r) + 2C(\tilde{t})\right) d\tilde{t}. \tag{2.49}$$

Here $C(t)$ is an arbitrary function of time, $M_{\rm T}$ is a constant, and a prime denotes a derivative with respect to the argument. The gauge function $C(t)$ results from the presence of a nonzero shift.

## E. Masses

In order to define the mass of a spherical body in Brans-Dicke theory, it is convenient to transform the solution (2.45)–(2.49) into a simpler gauge. Let

$$h^{\rm LT}_{ab} = h^{\rm MI}_{ab} + \zeta_{a,b} + \zeta_{b,a}, \tag{2.50}$$

where

$$\zeta_0 = \frac{1}{r}\int_0^t \left(M_{\rm T} + \frac{3}{2}f(\tilde{t}-r) + C(\tilde{t})\right) d\tilde{t}, \tag{2.51}$$

$$\zeta_i = 0. \tag{2.52}$$

Here $h^{\rm MI}_{ab}$ is the metric perturbation in maximal isotropic gauge, and $h^{\rm LT}_{ab}$ is the perturbation in Lorentz-Thorne (LT) gauge[16]. In LT gauge, we have

$$h^{\rm LT}_{00} = \frac{2M_{\rm T}}{r} + \frac{f(t-r)}{r}, \tag{2.53}$$

$$h^{\rm LT}_{0i} = 0, \tag{2.54}$$

$$h^{\rm LT}_{ij} = \delta_{ij}\left(\frac{2M_{\rm T}}{r} - \frac{f(t-r)}{r}\right), \tag{2.55}$$

$$\phi = 1 + \frac{f(t-r)}{r}. \qquad \text{(gauge invariant)} \tag{2.56}$$

Note that there is no shift in LT gauge.

For a static situation, $f$ is a constant, so it will appear as an additional "mass" in the metric. We will therefore write

$$f \equiv 2M_{\rm S} = \text{constant}, \qquad \text{time-independent case.} \tag{2.57}$$

Hence,

$$h^{\rm LT}_{00} = \frac{2M_{\rm T} + 2M_{\rm S}}{r}, \tag{2.58}$$

$$h^{\rm LT}_{0i} = 0, \tag{2.59}$$

$$h^{\rm LT}_{ij} = \delta_{ij}\left(\frac{2M_{\rm T} - 2M_{\rm S}}{r}\right), \tag{2.60}$$

$$\phi = 1 + \frac{2M_{\rm S}}{r}. \tag{2.61}$$

We see from Eq. (2.58) that a test particle in a Keplerian orbit measures a total mass $M$ equal to $M_{\rm T} + M_{\rm S}$. The "scalar mass" $M_{\rm S}$ is the portion of the active gravitational mass produced by the scalar field. As discussed further in Paper II, the "tensor mass" $M_{\rm T}$ is the mass measured by a test black hole in a Keplerian orbit. In general relativity, $M_{\rm S} = 0$ and $M_{\rm T} = M$.

Lee[17] has derived expressions and conservation laws for the tensor and scalar masses in terms of superpotentials. He defines the gauge-invariant quantities

$$\mathcal{M}_{\rm T} \equiv \frac{1}{16\pi} \int \left[\phi^2(-g)\left(g^{00}g^{ij} - g^{0i}g^{0j}\right)\right]_{,j} d^2\Sigma_i, \tag{2.62a}$$

$$\mathcal{M}_{\rm T} - \mathcal{M}_{\rm S} \equiv \frac{1}{16\pi} \int \left[(-g)\left(g^{00}g^{ij} - g^{0i}g^{0j}\right)\right]_{,j} d^2\Sigma_i, \tag{2.62b}$$

which in the time-independent case reduce to $M_{\rm T}$ and $M_{\rm T} - M_{\rm S}$, respectively. Here we assume Cartesian coordinates, and $\Sigma_i$ is the two-dimensional area element in the asymptotic rest frame of the source.

Evaluating Eqs. (2.62) using the metric (2.13), we obtain

$$\mathcal{M}_{\rm T} = \frac{1}{16\pi} \int \left(\phi^2 A^4\right)_{,r} r^2\, d\Omega = -\frac{r^2}{4}\left(\phi^2 A^4\right)_{,r}, \tag{2.63a}$$

$$\mathcal{M}_{\rm T} - \mathcal{M}_{\rm S} = \frac{1}{16\pi} \int \left(A^4\right)_{,r} r^2\, d\Omega = -\frac{r^2}{4}\left(A^4\right)_{,r}. \tag{2.63b}$$

In the linearized regime, we can expand these expressions to first order in $(\phi - 1)$ and $(A - 1)$. The result is

$$\mathcal{M}_{\rm T} = -r^2 A_{,r} - \frac{1}{2} r^2 \Phi, \tag{2.64a}$$

$$\mathcal{M}_{\rm S} = -\frac{1}{2} r^2 \Phi. \tag{2.64b}$$

Further simplification using the linearized equations (2.45)–(2.49) yields

$$\mathcal{M}_{\rm T} = M_{\rm T}, \tag{2.65a}$$

$$\mathcal{M}_{\rm S} = \frac{f}{2} + \frac{rf'}{2}. \tag{2.65b}$$

Although these reduce to $M_{\rm T}$ and $M_{\rm S}$ in the time-independent case, the scalar mass is not well-defined during dynamical epochs: $\mathcal{M}_{\rm S}$ approaches $rf'(t - r)/2$ as $r \to \infty$. The scalar mass has other unusual properties, which are discussed in detail by Lee[17]. For example, the scalar mass is not positive definite, and scalar mass carried by a gravitational wave does not curve up the background spacetime. Because the tensor mass $M_{\rm T}$ does not suffer from such difficulties, it is the quantity that most deserves to be called "mass". The quantity $M_{\rm T}$ is positive definite, can only decrease by the emission of gravitational radiation, and has other energy-like properties, unlike the scalar mass or the active gravitational mass $M$.

Gauge-invariant formulas for the fluxes of $M_{\rm T}$ and $M_{\rm S}$ can be expressed in terms of the Landau-Lifshitz pseudotensor and other pseudotensors involving the scalar field[17]. In the case of spherical symmetry, it is easier to obtain these expressions by differentiating Eqs. (2.63) with respect to time and using the field equations to eliminate second derivatives of the metric. The two methods must give the same answer. The result, to second order in the amplitudes, is

$$\begin{aligned}\frac{-2\mathcal{M}_{{\rm T},t}}{r^2} &= \frac{\beta}{r}\left(6A_{,r} + 3\Phi\right) - \frac{5}{2}\Phi K_{\rm T} + \Pi\Phi(\omega - 1)\\ &\quad - (\Pi + K_{\rm T})(4A_{,r} + \alpha_{,r}),\end{aligned} \tag{2.66a}$$

$$\begin{aligned}\frac{-2(\mathcal{M}_{\rm T} - \mathcal{M}_{\rm S})_{,t}}{r^2} &= \beta\left(\frac{6A_{,r}}{r} - \Phi_{,r} - \frac{2\Phi}{r}\right) + \Phi K_{\rm T} + \omega\Pi\Phi - K_{\rm T}(4A_{,r} + \alpha_{,r})\\ &\quad + \Pi_{,r}(1 + \epsilon - \xi + 4a),\end{aligned} \tag{2.66b}$$

$$\begin{aligned}\frac{-2\mathcal{M}_{{\rm S},t}}{r^2} &= \beta\left(\frac{5\Phi}{r} + \Phi_{,r}\right) - \frac{7}{2}\Phi K_{\rm T} - \Pi\Phi - \Pi(4A_{,r} + \alpha_{,r})\\ &\quad - \Pi_{,r}(1 + \epsilon - \xi + 4a),\end{aligned} \tag{2.66c}$$

where the variables $\epsilon$, $a$ and $\xi$ are defined by Eqs. (2.33)–(2.35), and we assume that the fluxes are evaluated in vacuum. Notice that $\mathcal{M}_{\mathrm{T},t} = 0$ to first order. This is because $M_{\mathrm{T}}$ is strictly constant in linearized theory. However, there is a nonzero first-order contribution to $\mathcal{M}_{\mathrm{S},t}$ (the $\Pi_{,r}$ term).

If we convert the surface integrals in Eqs. (2.63) to volume integrals by Gauss' theorem, and we eliminate second derivatives of $A$ using the Hamiltonian constraint, we obtain

$$\mathcal{M}_{\mathrm{T}} = \frac{1}{2}\int_0^\infty \left[8\pi\tilde{\rho}\phi A^3 + \frac{3}{4}A^6(K_{\mathrm{T}})^2\phi^2 + \left(\frac{\omega}{2} - 1\right)\Phi^2 A^4 \right. \\ \left. + \frac{\omega}{2}\Pi^2 A^6 - 7\phi\Phi A^3 A_{,r} - 7\phi^2 A^2 (A_{,r})^2\right] r^2\, dr, \tag{2.67a}$$

$$\mathcal{M}_{\mathrm{T}} - \mathcal{M}_{\mathrm{S}} = \frac{1}{2}\int_0^\infty \left[\frac{8\pi\tilde{\rho}A^3}{\phi} + \frac{3}{4}A^6(K_{\mathrm{T}})^2 + \frac{\omega A^4}{2\phi^2}\left(\Pi^2 A^2 + \Phi^2\right) \right. \\ \left. + \frac{\Phi}{\phi}A^3 A_{,r} - 7A^2(A_{,r})^2 + \frac{A^4\Phi_{,r}}{\phi} + \frac{2A^4\Phi}{\phi r}\right] r^2\, dr. \tag{2.67b}$$

These expressions will give us an additional check on our numerical code. Because the region of integration contains the origin, this check is only useful in the SA method.

## F. Light Rays and Apparent Horizons

From the metric (2.13), one can determine the coordinate velocity of radial light rays

$$\frac{dr}{dt}_{\pm} = \pm\frac{\alpha}{A} - \beta, \tag{2.68}$$

and the ingoing and outgoing radial null vectors

$$N_{\pm}^{a} = \left(1, \pm\frac{\alpha}{A} - \beta\right). \tag{2.69}$$

A marginally trapped surface is defined by

$$\nabla_{N_+}\mathcal{A} = 0, \tag{2.70}$$

where $\mathcal{A}$ is the area of a $(t, r = \mathrm{const})$ surface:

$$\mathcal{A} = 4\pi r^2 A^2. \tag{2.71}$$

In maximal isotropic gauge, Eq. (2.70) takes the form

$$\frac{1}{r} + \frac{A_{,r}}{A} = \frac{1}{2}AK_{\mathrm{T}}, \tag{2.72}$$

where we have used Eq. (2.19) to eliminate the time derivative of $A$.

The apparent horizon is the outermost marginally trapped surface. In our standard SA method, we use Eq. (2.72) to locate trapped surfaces. In the AHBC method, we place a grid point on the apparent horizon and use Eq. (2.72) as a boundary condition at that grid point.

For the AHBC method, we also need a relation that forces the apparent horizon to remain at a constant grid point as the metric variables and scalar field evolve in time. Such an equation can be obtained by differentiating Eq. (2.72) with respect to $t$ (holding $r$ constant), and then using Eqs. (2.19)–(2.22) to eliminate time derivatives, second radial derivatives of $A$, and first radial derivatives of $K_{\rm T}$. After some algebra, we arrive at the result

$$\frac{\alpha}{\beta A} = \frac{1 - 8\pi A^2 r^2 \left(\rho' + S_r'/A\right)}{1 + 8\pi A^2 r^2 \left({S^r}_r{}' + S_r'/A\right)}, \tag{2.73}$$

where the primed source terms are given by Eq. (2.32).

## III. SINGULARITY-AVOIDING (SA) METHOD

Here we present our singularity-avoiding numerical method for solving the Brans-Dicke equations for a spherically symmetric distribution of dust. Except for the addition of the Brans-Dicke source terms in the field equations, this method is very similar to that of Ref. [18].

We use a mean field particle simulation scheme in which the collisionless matter distribution is sampled by a finite number of noninteracting particles, and the gravitational field variables are defined on a finite number of grid zones. At each time step, the particles are moved according to the geodesic equation. By binning the particles, we calculate the source terms $\rho$, $S_r$, and $T$ in each grid zone. The source term $T$ is used to determine the scalar field via the wave equation, which is solved by finite differencing on the grid. All three source terms appear in the constraint equations, from which we determine the metric variables. The process is then repeated for the next time step.

### A. Matter Source Terms

We sample the matter by a finite number of noninteracting particles. If each particle has rest mass $m$, 4-velocity $u^a$, and comoving number density $\mathcal{N}$, the stress-energy tensor is given by

$$T^{ab} = \sum m\mathcal{N} u^a u^b, \tag{3.1}$$

where the sum is over all of the particles. The source terms defined by Eq. (2.26) are

$$\rho = \sum m\mathcal{N}(\alpha u^0)^2, \tag{3.2}$$

$$T = -\sum m\mathcal{N}, \tag{3.3}$$

$$S_r = -\sum m\mathcal{N}(\alpha u^0) u_r, \tag{3.4}$$

$${S^r}_r = \sum m\mathcal{N} u_r^2/A^2. \tag{3.5}$$

The comoving number density $\mathcal{N}$ of each particle is proportional to a delta function for point particles, but can be treated as a continuous quantity if the number of particles is large and the particle distribution is smooth. In evaluating the source terms, we average over all particles in each grid zone, so that we can think of each particle as occupying the entire volume of the zone. Therefore, we write

$$\frac{1}{\mathcal{N}} = 4\pi(\alpha u^0) A^3 r^2 \Delta r, \tag{3.6}$$

where $\Delta r$ is the width of a grid zone.

Notice that the source terms in Eqs. (3.2)–(3.5) depend on the metric function $A$. Since we use these source terms to solve for the metric functions, it is helpful to define quantities that can be calculated from the particle variables alone, without referring to the metric. This is especially important when solving the nonlinear Hamiltonian constraint (2.21) for $A$. We define

$$\tilde{\rho} \equiv \rho A^3 = \sum \frac{m(\alpha u^0)}{4\pi r^2 \Delta r}, \tag{3.7}$$

$$\tilde{T} \equiv T A^3 = -\sum \frac{m}{4\pi(\alpha u^0) r^2 \Delta r}, \tag{3.8}$$

$$\tilde{S}_r \equiv S_r A^3 = -\sum \frac{m u_r}{4\pi r^2 \Delta r}, \tag{3.9}$$

$$\tilde{S^r}_r \equiv {S^r}_r A^5 = \sum \frac{m u_r^2}{4\pi(\alpha u^0) r^2 \Delta r}. \tag{3.10}$$

Although the metric function $\alpha$ still appears in these expressions, it is always multiplied by $u^0$. The quantity $\alpha u^0$ depends only weakly on the metric (Eq. 3.12).

Simply evaluating the sums in Eqs. (3.7)–(3.10) in each grid zone gives reasonable values for the source terms in that zone. Even better is to use the algorithm of Hockney and Eastwood[19] to distribute each particle's rest mass among its own grid zone and each of the two neighboring zones. This procedure reduces the stochastic fluctuations associated with having only a finite number of particles, giving us smoother results[20].

## B. Geodesic Equations

Since the matter is made up of particles, the equation of motion $\nabla_a T^{ab} = 0$ reduces to the geodesic equation for each particle. Using the metric (2.13), we write the geodesic equation as

$$\frac{dr}{dt} = \frac{\alpha u_r}{A^2(\alpha u^0)} - \beta, \tag{3.11a}$$

$$\frac{du_r}{dt} = -(\alpha u^0)\alpha_{,r} + u_r \beta_{,r} + \frac{\alpha u_r^2}{(\alpha u^0)} \frac{A_{,r}}{A^3} + \frac{\alpha u_\phi^2}{(\alpha u^0) A^2 r^2} \left( \frac{1}{r} + \frac{A_{,r}}{A} \right). \tag{3.11b}$$

The normalization of four-velocity gives us

$$\alpha u^0 = \left( 1 + \frac{u_r^2}{A^2} + \frac{u_\phi^2}{A^2 r^2} \right)^{1/2}. \tag{3.12}$$

These equations are solved for the variables $r$, $u_r$, and $\alpha u^0$ for each particle at each time step by an embedded fourth-fifth order Runge-Kutta scheme with adaptive step size[21]. The quantity $u_\phi$ is a constant of the motion. Derivatives of the metric that appear in Eqs. (3.11)–(3.12) are obtained on the grid by finite differencing. The values of these derivatives and the metric functions at the particle position are then determined by interpolating from the grid.

## C. Scalar Wave Equation

After moving the particles and determining the source terms, we proceed to the wave equation. In order to minimize roundoff error in the nearly GR regime ($\phi$ close to unity), we use the variable

$$\xi \equiv \phi - 1. \tag{3.13}$$

The wave equation can then be written

$$\xi_{,t} - 2\beta r \xi_{,r^2} = -\alpha \Pi, \tag{3.14a}$$

$$\Pi_{,t} - 2\beta r \Pi_{,r^2} = \frac{8\pi \tilde{T}\alpha}{A^3(3+2\omega)} - \frac{6}{A^3}\left(r^3 A\alpha \xi_{,r^2}\right)_{,r^3}. \tag{3.14b}$$

These are Eqs. (2.29) and (2.31) written without explicit use of the variable $\Phi$. It is not necessary to use $\Phi$ as a dynamical variable until we introduce the AHBC method in Section IV.

The regularity condition at the origin is

$$\xi_{,r} = \Pi_{,r} = 0. \tag{3.15}$$

Since $\xi$ and $\Pi$ behave like

$$C_1 + C_2 r^2, \qquad C_1, C_2 \quad \text{constant} \tag{3.16}$$

for small $r$, we take radial derivatives with respect to $r^2$ rather than $r$ in Eqs. (3.14). This ensures that our finite difference equations yield the correct result near the origin[22].

The outgoing-wave condition at infinity is

$$(rY)_{,r} + (rY)_{,t} = 0, \tag{3.17}$$

where $Y$ is either $\Pi$ or $\xi$. For spherical symmetry, Eq. (3.17) is exact in linearized theory, as one can verify by substituting Eq. (2.45).

We solve Eqs. (3.14) by an explicit staggered leapfrog scheme that determines $\xi$ and $\Pi$ at timestep $n+1$ given their values at timesteps $n$ and $n-1$. No information at timestep $n+1$ is needed to solve the equations. After determining $\xi$ and $\Pi$, we calculate $\phi$ from Eq. (3.13). Finite difference approximations are presented in Appendix A.1.

## D. Constraints

After determining the quantities $\xi$, $\Pi$, and $\phi$, we then solve for $A$ and $K_{\mathrm{T}}$.

By introducing the new variables

$$Z \equiv A^3 r^3 \phi K_{\mathrm{T}}, \tag{3.18}$$

$$\psi \equiv A^{1/2}, \tag{3.19}$$

we can rewrite the momentum constraint (2.22) and the Hamiltonian constraint (2.21) in the form

$$5Z_{,r^5} = \frac{8\pi \tilde{S}_r}{r} - 2\psi^6 \Pi_{,r^2} - \frac{2\psi^6 \omega \Pi \xi_{,r^2}}{\phi}, \tag{3.20}$$

$$\frac{6}{\phi^{1/2}}\left(r^3 \phi^{1/2} \psi_{,r^2}\right)_{,r^3} = -\frac{3}{16 r^6 \psi^7 \phi^2} Z^2 - \frac{2\pi\tilde{\rho}}{\phi\psi} - \frac{\omega\psi^5 \Pi^2}{8\phi^2} - \frac{\omega\psi}{2}\left(\frac{r\xi_{,r^2}}{\phi}\right)^2 - \frac{3\psi}{2}\left(r^3 \xi_{,r^2}\right)_{,r^3}. \tag{3.21}$$

In general relativity $\Pi = \xi_{,r} = 0$, so one can determine $Z$ from Eq. (3.20) and then compute $\psi$ from Eq. (3.21). This is the reason for using the variable $Z$ instead of $K_{\mathrm{T}}$. In Brans-Dicke theory, Eqs. (3.20) and (3.21) are coupled because of the last two terms in Eq. (3.20), so we must solve these equations simultaneously.

Eq. (3.20) requires a single boundary condition. For regularity, we set $Z = 0$ at the origin. Eq. (3.21) requires two boundary conditions. The regularity condition at the origin is

$$\psi_{,r} = 0. \tag{3.22}$$

At infinity, we match to the linearized solution (2.46). Differentiating this equation yields

$$(r\psi)_{,r} = 1 - \frac{r\Pi}{4}, \tag{3.23}$$

which does not require knowledge of the tensor mass $M_{\mathrm{T}}$.

We determine the variable $Z$ using the momentum constraint (3.20), and we solve the nonlinear Hamiltonian constraint (3.21) for $\psi$ using the iterative scheme described in Appendix A.2. Because $\psi$ appears in Eq. (3.20), we must recompute $Z$ at each step in the iteration. We obtain an initial guess for $\psi$ from the evolution equation (2.19), which we write as

$$\psi_{,t} - \beta\psi_{,r} = \frac{\beta\psi}{2r} - \frac{1}{4}\alpha\psi K_{\mathrm{T}}. \tag{3.24}$$

After determining $Z$ and $\psi$, the variables $A$ and $K_{\mathrm{T}}$ are found from Eqs. (3.18) and (3.19). Finite difference forms of Eqs. (3.20)–(3.24) are presented in Appendix A.2.

## E. Lapse and Shift

Having determined the scalar field variables and the spatial metric variables, we can now calculate $\alpha$ and $\beta$. The lapse equation (2.23) can be written

$$\frac{6}{\alpha A^3}\left(r^3 A\alpha_{,r^2}\right)_{,r^3} = \frac{6}{\phi A^3}\left(r^3 A\xi_{,r^2}\right)_{,r^3} + \frac{8\pi}{\phi}\left(\tilde{\rho} + \frac{\tilde{T}}{2+3/\omega}\right) + \frac{\omega\Pi^2}{\phi^2} + \frac{3}{2}K_{\mathrm{T}}^2. \tag{3.25}$$

The regularity condition at the origin is

$$\alpha_{,r} = 0. \tag{3.26}$$

We match to the linearized solution to obtain the boundary condition at infinity: by differentiating Eq. (2.47), we obtain

$$(r\alpha)_{,r} = 1 + r\Pi. \tag{3.27}$$

This condition is independent of the gauge function $C(t)$ that appears in Eq. (2.47).

After solving for $\alpha$, we calculate $\beta$ from the shift equation (2.24). This equation requires a single boundary condition. We impose Eq. (2.48) at the outermost grid point:

$$\beta = \frac{rK_{\mathrm{T}}}{2} + \frac{r\Pi}{2}. \tag{3.28}$$

Finite difference forms of Eqs. (3.25)– (3.28) are given in Appendix A.3.

## F. Grid

Our numerical grid extends from $r_1 = 0$ to $r_{i_{\mathrm{max}}} = r_{\mathrm{max}}$. All variables are centered at half-grid points $r_{i+1/2}$. In order to obtain a nearly constant number of particles in each grid zone, we divide the grid into inner and outer regions. The inner region, which contains all the particles, extends from $r_1$ to $r_{i_{\mathrm{part}}}$ where $r_{i_{\rm part}}$ is the grid point just outside the outermost particle. The grid point $r_2$ is chosen so that it contains a fraction $1/i_{\rm part}$ of the rest mass. Spacing between other grid points in the inner region is geometric in $r^3$:

$$\frac{r_{i+1}^3 - r_i^3}{r_i^3 - r_{i-1}^3} = \frac{r_{i+2}^3 - r_i^3}{r_{i+1}^3 - r_i^3}. \tag{3.29}$$

This ensures that the grid spacing is close to uniform in $r^3$, so that for a uniform particle distribution each zone has approximately the same number of particles. The outer grid, which extends from $r_{i_{\rm part}}$ to $r_{i_{\rm max}}$, is matched smoothly onto the inner grid. The spacing between outer grid points is also geometric in $r^3$.

The grid is allowed to move at every time step so that it follows the particle distribution. Each time the grid is moved, all gravitational field variables are interpolated from the old grid onto the new one. Moving the grid allows us to place many grid points where they are required to maintain accuracy. If the grid remained stationary during gravitational collapse, the particles would soon end up only in the innermost grid zones, invalidating our finite difference approximations.

### G. Identifying Apparent Horizons

To determine the location of an apparent horizon, we evaluate the quantity

$$\theta = \frac{1}{r} + 2\frac{\psi_{,r}}{\psi} - \frac{1}{2}\psi^2 K_{\rm T} \tag{3.30}$$

at each grid point. By Eq. (2.72), all grid points with $\theta < 0$ are contained within a trapped surface. Therefore, if $r_{i_{\rm AH}}$ is the outermost grid point with $\theta < 0$, then the apparent horizon lies between $r_{i_{\rm AH}}$ and $r_{i_{\rm AH}+1}$. We obtain the approximate value of $r_{\rm AH}$ by

$$r_{\rm AH} = r_{i_{\rm AH}} - \theta_{i_{\rm AH}} \frac{r_{i_{\rm AH}+1} - r_{i_{\rm AH}}}{\theta_{i_{\rm AH}+1} - \theta_{i_{\rm AH}}}. \tag{3.31}$$

### H. Initial Data

Our initial time slice occurs at a moment of time symmetry, so that $K_{\rm T} = \Pi = \beta = 0$. We first calculate the initial values of the metric and scalar field using a method similar to that of Matsuda[23]. This method is discussed in Appendix B. It involves only ODEs, which can be solved to arbitrary accuracy by standard numerical methods.

After obtaining the ODE solution, we re-solve for the initial data using the mean-field particle scheme. We first randomly place particles in the interior according to the rest mass function $M_{\rm rest}(r)$ obtained from the ODE solution. We then compute the matter source terms by binning the particles, and we solve for $\xi$ using Eq. (3.14b ) with $\Pi = \beta = 0$. Next we solve for $\psi$ and $\alpha$ using the Hamiltonian constraint (3.21) and the lapse equation (3.25). By comparing $\xi$, $\psi$, and $\alpha$ with the result of the ODE solution, we obtain an important check on our method.

### I. Diagnostics

There are two evolution equations, Eqs. (2.19) and (2.20), that we did not use in solving the Brans-Dicke equations (Although we used Eq. (2.19) as an initial guess for $\psi$, we refined our guess using the Hamiltonian constraint). We can test the accuracy of our code by comparing the left and right-hand sides of these equations at each time step.

Another check is obtained by mass conservation. In the linearized region far from the origin, we calculate the quantities $\mathcal{M}_S$ and $\mathcal{M}_T$ by Eqs. (2.65), and compare with the result given by Eqs. (2.66) and with the volume integrals (2.67). This is more useful than evaluating the evolution equations at each time step because it is sensitive to errors that accumulate over time.

## IV. APPARENT HORIZON BOUNDARY CONDITION (AHBC) METHOD

Here we present our AHBC method of solving the Brans-Dicke equations for a spacetime that contains a black hole. Our method depends on two primary ingredients. The first is the existence of what we call a coordinate causal horizon (CCH), which is a coordinate surface inside of which the coordinate velocity of radially outgoing light rays is negative, so that instantaneously any point inside the CCH cannot causally influence a point in the exterior. Note that a CCH is coordinate dependent, and can exist even in Minkowski space by choosing a coordinate system that moves faster than light.

The second ingredient is a horizon-locked coordinate system, similar to that of Seidel and Suen[9]. We place the AH at a fixed radial coordinate $r_{\rm AH}$, and require it to remain there at all times.

We truncate our spacetime at some radial coordinate $r_1$ such that $0 < r_1 \leq r_{\rm AH}$, and we discard the interior region. As long as a CCH exists for some $r \geq r_1$, discarding the interior cannot in principle affect the evolution of the exterior. In general relativity, a horizon-locked coordinate system guarantees that the AH is on or inside a CCH, since a light ray at the AH cannot propagate outwards. In this case, one can set $r_1 = r_{\rm AH}$. In Brans-Dicke theory, we find that this is not the case, as discussed further in Paper II. We therefore leave a small buffer zone between $r_1$ and $r_{\rm AH}$, and check that a CCH is always present in the portion of spacetime that we retain. Even in general relativity, this buffer zone is also useful for computing radial derivatives at the AH.

We solve the wave equation in a manner that takes advantage of the causal structure of the spacetime. We define a "causal boundary" at some radial grid point $r_{\rm CB}$ which coincides with either the AH or the CCH, whichever is smaller. Given any $r' \leq r_{\rm CB}$, our difference scheme ensures that the scalar field at $r'$ depends only on quantities at $r > r'$. Furthermore, no explicit boundary condition is needed at the inner grid point $r_1$.

To obtain the metric variables $\alpha$, $A$, $\beta$, and $K_{\rm T}$, we solve spatial constraint equations that require a total of six boundary conditions. Three of these are provided by matching to the linearized solution at the outer grid point. The other three, in the SA method described in the last section, were obtained by regularity at the origin. In the AHBC method, however, we exclude the origin from our spacetime, so these three boundary conditions must be imposed at the AH. We obtain two conditions by locking the AH to a fixed radial grid point. One is the marginally trapped surface relation (2.72), and the other, Eq. (2.73), is the requirement that the AH remain at a constant coordinate radius for all time. A third condition could be obtained by setting the tensor mass, $M_{\rm T}$, of the black hole to the value obtained by mass conservation, Eq. (2.66a ). However, this would prevent us from using mass conservation as a check on our numerical integrations. Instead, we use a different approach: we use the evolution equation (2.19) to update $A$ on the AH. Another possibility would be to evolve the value of $K_{\rm T}$ on the AH using Eq. (2.20), but this would be more complicated.

### A. Matter

As in the SA method, the particles can be moved according to the geodesic equations and the matter source terms can be computed by binning particles into zones. However, in the case of gravitational collapse starting with a uniform density profile, we will find that an AH does not form until after all matter has fallen into the black hole. For this reason, we only need to solve the AHBC equations in vacuum—it is not necessary to include matter particles in the code. This is a great advantage in terms of efficiency, since in the SA scheme most of the computer time is spent moving the particles, and it also gives us flexibility in the grid choice, since we no longer need to choose grid zones that are approximately constant in volume.

We emphasize that there is no fundamental restriction that prevents us from including matter in the AHBC method, and for some physical situations not investigated in this paper, *e.g.,* the collapse of a distribution with a core-halo structure, an AH may form while there is still matter in the exterior region. In such a case, including particles in the code would not be very difficult. For this reason, in the following sections we will continue to include the matter source terms in our equations.

### B. Grid

For accuracy, it is usually desirable to choose the finite difference grid to be uniformly spaced. However, a grid uniformly spaced in $r$ does not yield enough grid coverage near the AH. This is because the initial data for the AHBC method is determined by matching onto the SA method after a horizon is formed, and the SA method typically produces an apparent horizon radius $r_{\rm AH}$ much less than the outer grid radius and metric quantities that vary exponentially with $r$ near the AH. It is therefore convenient to choose a logarithmically spaced grid, and to take all radial derivatives with respect to $\ln r$ rather than $r$. This choice places many grid points near the AH where they are needed, and still allows us to take finite differences on a uniformly spaced (in $\ln r$) grid.

For convenience, we define the new variable

$$\eta \equiv \ln r. \tag{4.1}$$

Our radial grid then consists of the $i_{\rm max}$ points $(\eta_1, \eta_2, \ldots, \eta_{i_{\rm AH}}, \ldots, \eta_{i_{\rm max}})$. The AH is located at the grid point $i = i_{\rm AH}$. The point $r_{i_{\rm max}}$ is chosen to be far from the black hole, in a region where linearized theory is valid. All variables except $\Phi$ are defined on the grid points; $\Phi$ is defined on the half-grid points $(\eta_{3/2}, \ldots, \eta_{i_{\rm max}+1/2})$. The staggering of $\Phi$ is essential for our method of solving the wave equation.

Unfortunately, spacing the grid uniformly in $\ln r$ tends to produce too sparse a grid at large radii. A consequence of this is that outgoing waves get partially reflected at the outer edge of the grid. To minimize this effect, we pick a maximum threshold $\Delta r_0$ for the spacing between grid points. We choose our grid uniform in $\eta$ unless this choice would yield a grid spacing (in $r$) greater than $\Delta r_0$ at the outer boundary. In the latter case, we set $r_{i_{\rm max}} - r_{i_{\rm max}-1}$ equal to $\Delta r_0$, and choose the other grid points so that

$$\frac{\eta_{i+1} - \eta_i}{\eta_i - \eta_{i-1}} = \frac{\eta_{i+2} - \eta_{i+1}}{\eta_{i+1} - \eta_i}. \tag{4.2}$$

This gives us a grid that is as close as possible to being uniform in $\eta$ while still obeying $r_{i+1} - r_i \leq \Delta r_0$ for all $i$.

### C. Wave equation

The wave equation can be written in the form

$$\xi_{,t} = \beta\Phi - \alpha\Pi, \tag{4.3}$$

$$\Phi_{,t} = \frac{1}{r}\left[\beta\Phi_{,\eta} + \beta_{,\eta}\Phi - (\alpha\Pi)_{,\eta}\right], \tag{4.4}$$

$$\Pi_{,t} = \frac{\beta}{r}\Pi_{,\eta} + \frac{8\pi T\alpha}{3+2\omega} - \frac{\alpha}{\psi^4 r}\left[\Phi_{,\eta} + \Phi\left(2 + \frac{\alpha_{,\eta}}{\alpha} + 2\frac{\psi_{,\eta}}{\psi}\right)\right]. \tag{4.5}$$

We solve Eqs. (4.4) and (4.5) for $\Pi$ and $\Phi$ by the causal method described in Appendix C. Because this method takes into account the causal structure of the spacetime, it is not necessary to impose explicit boundary conditions at the inner edge of the grid. Boundary conditions at the outer edge of the grid are obtained by matching both $\Pi$ and $\Phi$ to the outgoing-wave linearized solution. Eqs. (2.37), (2.38), and (2.45) yield

$$\Pi_{,t} + \frac{1}{r}\Phi_{,\eta} + \frac{2\Phi}{r} = 0, \tag{4.6}$$

$$\Phi_{,t} + \frac{1}{r}\Phi_{,\eta} + \frac{2\Phi}{r} - \frac{\Pi}{r} = 0. \tag{4.7}$$

After determining $\Phi$ and $\Pi$, we solve for $\xi$ using Eq. (4.3). This equation is an ODE in time and requires no boundary conditions. The scalar field $\phi$ is then calculated from Eq. (3.13).

### D. Constraints

After solving the wave equation, the Hamiltonian and momentum constraints are solved simultaneously for the variables $\psi$ and $Z$, from which we can compute $A$ and $K_{\rm T}$. These equations are coupled because scalar wave terms containing $\psi$ appear in the momentum constraint (2.22). Eqs. (2.21) and (2.22) can be written as

$$Z_{,\eta} = 8\pi r^4 \tilde{S}_r - \psi^6 r^3 \left(\Pi_{,\eta} + \frac{\Pi\Phi\omega r}{\phi}\right), \tag{4.8}$$

$$\begin{aligned}\psi_{,\eta\eta} + \psi_{,\eta}\left(1 + \frac{\Phi r}{2\phi}\right) = &- \frac{3}{16}\frac{Z^2}{\phi^2\psi^7 r^4} - \frac{2\pi\tilde{\rho} r^2}{\phi\psi} - \frac{\omega\psi^5\Pi^2 r^2}{8\phi^2} \\ &- \frac{\omega\psi\Phi^2 r^2}{8\phi^2} - \frac{r\psi}{4\phi}\left(\Phi_{,\eta} + 2\Phi\right).\end{aligned} \tag{4.9}$$

Eq. (4.8) requires a single boundary condition. At the apparent horizon, we impose the marginally trapped surface condition (2.72), which we rewrite using the variables $\psi$, $Z$, and $\eta$:

$$\psi_{,\eta} + \frac{\psi}{2} = \frac{1}{4}\frac{Z}{\psi^3 r^2 \phi}. \tag{4.10}$$

Eq. (4.9) requires two boundary conditions. At the outermost grid point, we match to linearized theory by imposing Eq. (3.23):

$$(r\psi)_{,\eta} = r - \frac{r^2\Pi}{4}. \tag{4.11}$$

Because there is no extra equation for the inner boundary condition, we use the evolution equation (2.19) to compute the value of $\psi$ at the apparent horizon. This equation takes the form

$$\frac{\psi_{,t}}{\psi} = \beta\frac{\psi_{,\eta}}{\psi r} + \frac{\beta}{2r} - \frac{1}{4}\alpha K_{\rm T}. \tag{4.12}$$

Eqs. (4.8)—(4.11) comprise a coupled system of nonlinear equations. We solve them simultaneously by the iteration scheme described in Appendix D.

### E. Lapse and Shift

After obtaining $\psi$ and $Z$, we solve the lapse and shift equations simultaneously for $\alpha$ and $\beta$. In the SA method, these equations are solved independently; here they are coupled because of a boundary condition that we now impose at the apparent horizon. We write these equations as

$$\alpha_{,\eta\eta} + \alpha_{,\eta}\left(1 + 2\frac{\psi_{,\eta}}{\psi}\right) = \alpha\left[\frac{3}{2}\left(K_{\rm T} r\psi^2\right)^2 + \frac{8\pi r^2}{\phi\psi^2}\left(\frac{\tilde{\rho}+\tilde{T}}{2+3/\omega}\right)\right.$$
$$\left. + \frac{\omega\psi^4\Pi^2 r^2}{\phi^2} + \frac{r}{\phi}\left(\Phi_{,\eta} + 2\Phi\left(1 + \frac{\psi_{,\eta}}{\psi}\right)\right)\right], \tag{4.13}$$

$$\left(\frac{\beta}{r}\right)_{,\eta} = -\frac{3}{2}\alpha K_{\rm T}. \tag{4.14}$$

At the outer grid point, we impose Eqs. (3.27) and (2.48), which take the form

$$(r\alpha)_{,\eta} = r + r^2\Pi, \tag{4.15}$$

$$\beta = \frac{rK_{\rm T}}{2} + \frac{r\Pi}{2}. \tag{4.16}$$

For the inner boundary condition on $\alpha$ we use Eq. (2.73), which can be written as

$$\frac{\alpha}{\beta\psi^2} = \frac{1/r^2 - F_1 - F_3}{1/r^2 + F_2 + F_3 + F_4}, \tag{4.17}$$

where

$$F_1 \equiv \frac{8\pi}{\phi}\left(\tilde{\rho} + \frac{\tilde{S}_r}{\psi^2}\right), \tag{4.18a}$$

$$F_2 \equiv \frac{8\pi}{\phi}\left(\tilde{S}^r{}_r - \frac{\tilde{T}}{3+2\omega} + \frac{\tilde{S}_r}{\psi^2}\right), \tag{4.18b}$$

$$F_3 \equiv \frac{\omega}{2\phi^2}\left(\Pi\psi^2 - \Phi\right)^2 - \frac{\psi^2\Pi_{,\eta}}{r\phi} + \frac{\Phi_{,\eta}}{r\phi} - 2\frac{\Phi\psi_{,\eta}}{r\phi\psi}, \tag{4.18c}$$

$$F_4 \equiv \frac{K_{\rm T}A}{\phi}\left(A\Pi - \Phi\right). \tag{4.18d}$$

This condition forces a grid point to remain at the apparent horizon. It also couples the lapse and shift equations (4.13) and (4.14), so that we must solve these equations simultaneously. Finite-difference forms of Eqs. (4.14)–(4.18) are presented in Appendix D.

Notice that in vacuum with constant scalar field, Eq. (4.17) reduces to $\alpha = A\beta$. Combining this with Eqs. (4.10) and (4.12), we see that in this case $\psi_{,t} = 0$ on the apparent horizon. As a result, *all* quantities are *manifestly time independent:* Given the value of $\psi$ on the AH, which is constant in time, the quantities $A$, $\alpha$, $\beta$, and $K_{\rm T}$ are completely determined by coupled spatial constraint equations with no source terms.

This time independence should come as no great surprise because the vacuum solution with constant scalar field is simply the exterior Schwarzschild solution. However, this result is remarkable from a practical point of view: When one uses standard singularity-avoiding schemes (including our SA scheme) to compute spherical collapse in general relativity, one does not obtain a time-independent system at the end of the simulation. Instead, one finds that the metric functions and their derivatives are changing exponentially with time inside the AH. However, using the AHBC method, any system that results in a Schwarzschild black hole will become manifestly time independent. This is the great benefit of the AHBC method: One can integrate black-hole spacetimes arbitrarily far into the future without encountering singularities and without causing metric functions or their derivatives to blow up.

## F. Initial Data

The initial time slice for the AHBC method can be any maximal isotropic time slice that contains an apparent horizon. We use a slice provided by the SA method. Because $\phi$ is a scalar and $\psi$, $K_{\rm T}$, and $\Pi$ are three-dimensional geometric objects on the slice, and because we use the same 3-metric for both the SA and AHBC methods, the values of these variables on the initial slice can be simply read off from the SA slice. After laying down a new spatial grid as described in section A, we spatially interpolate these variables from the old SA grid to the new one. We then determine $\Phi$ from Eq. (2.28). For self-consistency, we then freeze the value of $\psi$ at the AH and recalculate $\psi$ and $K_{\rm T}$ from Eqs. (4.8)–(4.11). Finally, we calculate $\alpha$ and $\beta$ by solving Eqs. (4.14)–(4.18).

Although the initial AHBC time slice coincides with an SA slice, subsequent AHBC slices do not coincide with those that would have been produced by continuing the SA method, because the two methods have a different lapse and shift. These differences are due soley to the different inner boundary conditions used in the two methods.

### G. Diagnostics

In determining the metric and the scalar field, we never use the $K_{\rm T}$ evolution equation (2.20) or the definition of $\Phi$, Eq. (2.28). Furthermore, while we need the $A$ evolution equation (2.19) to determine a boundary condition at the AH, this equation is not used at any other grid point except as an initial guess. We therefore have three equations that can serve as diagnostics at each time step.

In addition, we can compare the results of the masses calculated by Eqs. (2.65) and (2.66). Unlike the step-by-step comparison of equations (2.20), (2.28), and (2.19), this method is sensitive to errors that accumulate over time. Because we exclude the origin from our spacetime, we cannot evaluate the volume integrals (2.67) in the AHBC method.

## V. NUMERICAL RESULTS

In this section we treat a few select collapse scenarios to calibrate our code. We use these cases to compare the standard SA scheme with the AHBC method. For a more complete summary of physical results and an evaluation of collapse to black holes in Brans-Dicke theory, see Paper II.

### A. Oppenheimer-Snyder Collapse in General Relativity

A useful test of our code is Oppenheimer-Snyder[24] collapse to a black hole in general relativity. We start with a momentarily static, uniform sphere of dust with an initial areal radius $R_s = 10M$, where $M$ is the total mass of the configuration as measured by Keplerian orbits far from the origin. We follow the collapse of this object to a black hole using our SA scheme with $\omega = 10^{37}$. Such a large value of $\omega$ puts us well into the general relativistic limit. We use 41 interior grid points, 87 exterior grid points, and 1200 particles. Our outer grid boundary is at $r = 100M$.

Figures 1, 2, and 3 show snapshots of metric parameters at selected values of coordinate time. Also shown are the exact results obtained[25] by transforming the analytic solution[24] into the maximal isotropic gauge. The SA method agrees well with the exact solution until very late times. The "collapse of the lapse" can be seen in Figure 1: inside $r_s = 1.5M$, the lapse function decreases exponentially with time, freezing the clocks of normal observers in this region. The values of $A$ and $\beta$ inside $r_s = 1.5M$ also change rapidly with time, as can be seen in Figures 2 and 3. Because of the large values of $A$, coordinate radii become very small compared to isotropic radii. For example, the areal radius of the matter surface at $t = 87M$ is about $r_s = 1.5M$, while its coordinate radius is only $r \sim 5 \times 10^{-8}M$. In addition, the gradients of $\alpha$, $A$, and $\beta$ become very steep near $r_s = 1.5M$.

Tracking particles at $r \sim 10^{-8}M$ with a grid that extends to $r = 100M$ and resolving the metric function $A$ which drops from $> 10^7$ to 2 between $r = 10^{-7}M$ and $r = M$ requires an enormous dynamic range.

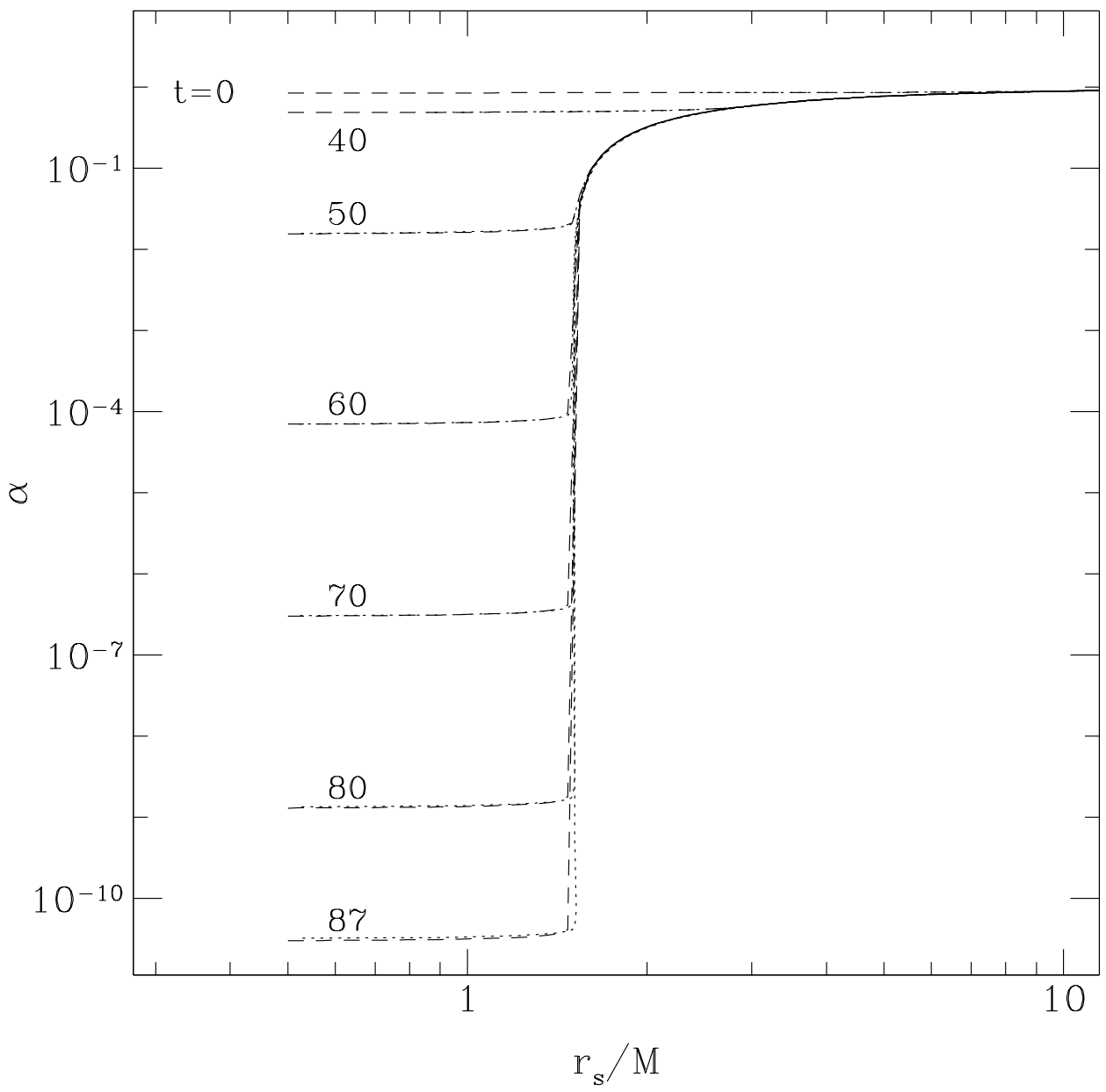

**Figure 1** The lapse function $\alpha$ versus areal radius $r_s$ on selected maximal time slices for general relativistic Oppenheimer-Snyder collapse from $R_s = 10M$. Calculations by the SA method (dotted line) are compared with the exact solution (dashed line). Time and distance are measured in units of $M$.

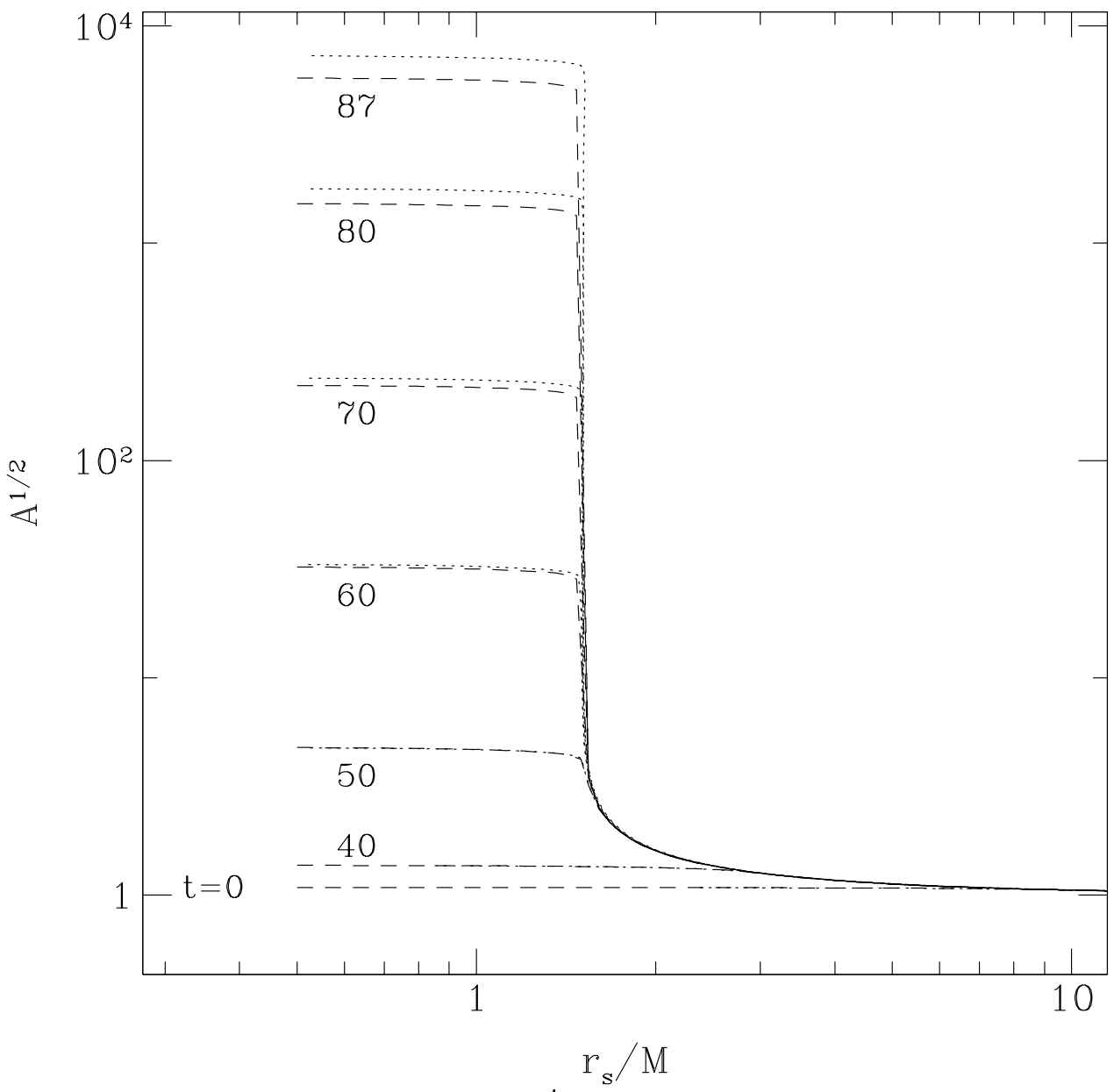

**Figure 2** The metric component $A^{1/2}$ versus areal radius $r_s$ on selected maximal time slices. Labeling is the same as in Figure 1.

Because our inner grid follows the particles at each time step, grid points tend to accumulate at the limit (matter) surface, $r_s \sim 1.5M$. Although this allows us to resolve the large metric gradients better than with a stationary grid, our code eventually loses accuracy as the gradients grow. This can be seen in Figures 1–3:

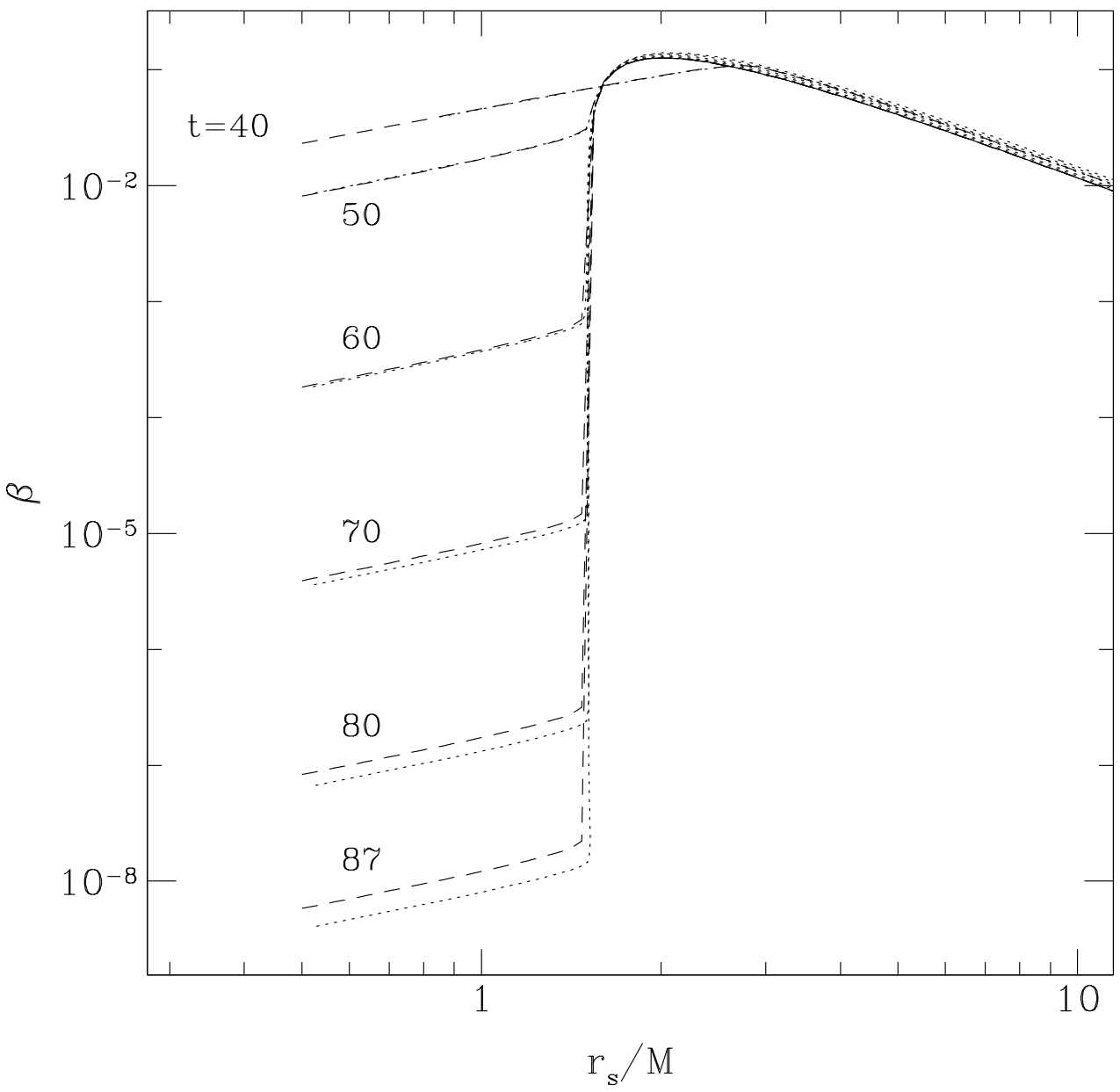

**Figure 3** The radial shift $\beta$ versus areal radius $r_s$ on selected maximal time slices. Labeling is the same as in Figure 1.

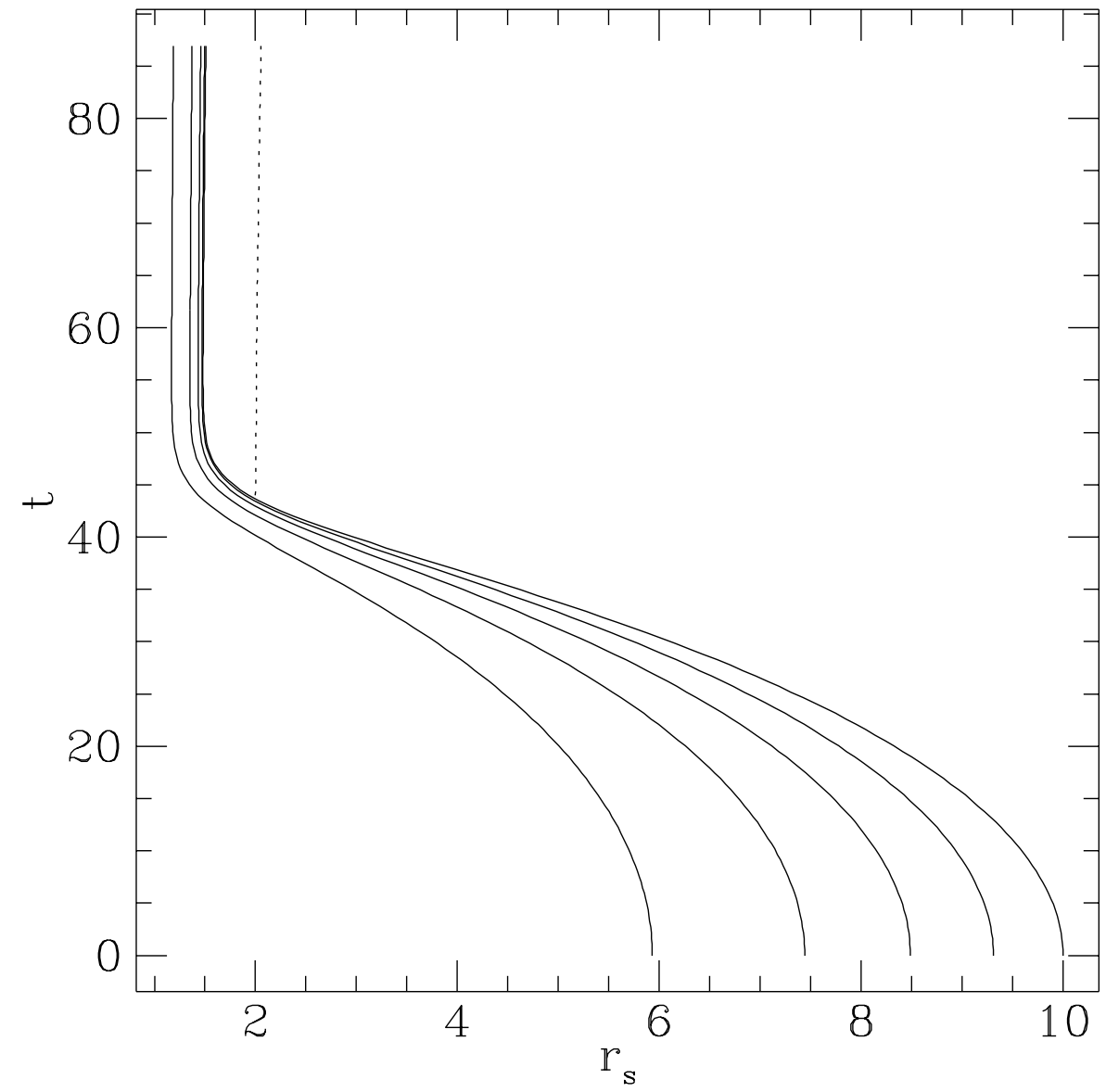

**Figure 4** Oppenheimer-Snyder collapse in maximal slicing, calculated by the SA method. The five solid lines represent world lines of matter elements containing 20%, 40%, 60%, 80%, and 100% of the interior rest mass. Time and areal radius are measured in units of $M$. The dotted line is the apparent horizon, which forms at about $t = 44M$, at which point it already coincides with the event horizon.

the SA method and the exact soultion begin to disagree at about $t = 60M$, and this discrepancy increases exponentially with time. Even if we were able to maintain accuracy, numerical overflow would eventually cause our SA code to terminate.

A spacetime diagram of Oppenheimer-Snyder collapse is shown in Figure 4. Because of the "collapse of the lapse", the matter particles sit at a constant areal radius after $t = 50M$. An apparent horizon, which forms at $t = 44M$ outside of the matter, has an initial areal radius of $r_s = 2M$, in agreement with the Schwarzschild solution. It increases slightly in size after about $t = 70M$ because of numerical errors.

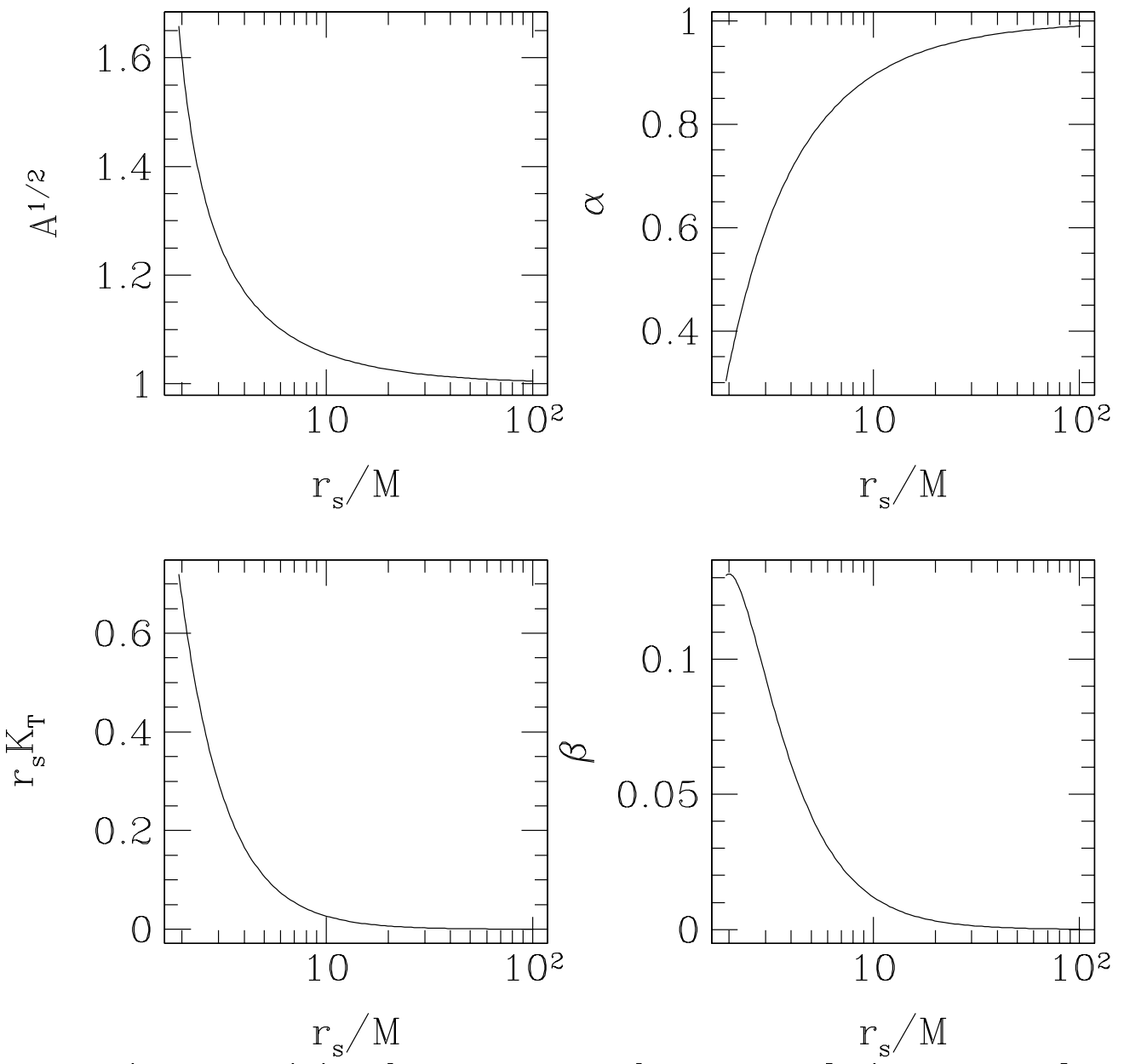

**Figure 5** Metric quantities for $t > 45M$ for Oppenheimer-Snyder collapse in general relativity, calculated by the AHBC method. We remove the region of spacetime inside the apparent horizon at $r_s = 2M$ and only retain the vacuum exterior. All quantities are constant in time.

The above numerical difficulties do not occur in the AHBC method because we no longer use such a pathological coordinate system. In Figure 5 we show the metric coefficients obtained from the AHBC method after matching onto an SA time slice at $t = 45$. We use 128 grid points, and place the outer boundary at $r = 100M$. The apparent horizon is located at the grid point $i = 4$, which is at coordinate radius $r = 0.78M$ and areal radius $r_s = 2M$. The innermost grid point is at $r = 0.7M_0$. Because all the matter has fallen past the AH by $t = 45M$, we no longer need to move the particles—we have discarded the region of spacetime that contains them. In this case the metric is manifestly static.

B. Scalar Waves on a Schwarzschild Background

Consider a small scalar perturbation about a Schwarzschild black hole. If the perturbation is so small that its contribution to the metric is negligible, then it is not necessary to solve the coupled Brans-Dicke equations (2.4) and (2.5) to determine the future evolution of $\phi$. Instead, one only needs to solve the wave equation (2.4) in vacuum on a Schwarzschild background. This equation can be written in the form[26, 27]

$$u_{,t_s t_s} = u_{,zz} + Vu, \tag{5.1}$$

where

$$u \equiv r_s \phi, \tag{5.2}$$

$$V \equiv \frac{2}{r_s^3}\left(1 - \frac{2}{r_s}\right), \tag{5.3}$$

$$z \equiv r_s + 2\ln(r_s/2 - 1), \tag{5.4}$$

$r_s$ is the areal radius, and $t_s$ is the Schwarzschild time coordinate. The variable $z$ is the familiar "tortoise" coordinate, which runs from $z = -\infty$ (at $r_s = 2$) to $z = \infty$ (at $r_s = \infty$). The mass of the black hole has been set equal to unity. Eq. (5.1) provides an important check for our AHBC code. This equation is not difficult to solve numerically—it is simply a one-dimensional flat space wave equation in a static potential.

We set up the following test case: At $t_s = 0$, the metric is Schwarzschild and the scalar field is given by

$$\phi = 1 + \frac{C}{r_s}\exp\left(\frac{(r_s - r_0)^2}{2\sigma^2}\right), \tag{5.5}$$

where $r_0 = 80$, $C = 10^{-6}$, and $\sigma = 5$. We set $\phi_{,t_s} = 0$ initially. As time progresses, the initial pulse divides into two pieces: one moves outwards to infinity, while the other moves toward the black hole and is partially reflected by the Schwarzschild potential.

We calculate $\phi$ as a function of $r_s$ and $t_s$ for this case by two independent methods: we solve the full Brans-Dicke equations using the AHBC method, and we solve the perturbation equation (5.1) using a staggered leapfrog finite difference scheme. The AHBC method requires metric coefficients on an initial time slice; these are provided by matching onto Oppenheimer-Snyder collapse (calculated from the exact solution following Petrich *et al.*[25]) after the matter has fallen into the black hole.

Because the AHBC method and the perturbation method use different coordinate systems, we must be careful in setting up the initial data. Eq. (5.5) is defined on the Schwarzschild time slice $t_s = 0$. This slice does not correspond to the maximal slice $t = 0$, or any other $t =$ constant slice in maximal isotropic gauge. However, the initial Gaussian pulse (5.5) is far from the black hole, where $t =$ constant and $t_s =$ constant slices nearly coincide. Therefore, if we choose the slice $t = 0$ to coincide with $t_s = 0$ at $r_s = 80$, we can use Eq. (5.5) to determine $\phi$ at $t = 0$ for the AHBC method without introducing much error.

Although $t =$ constant and $t_s =$ constant slices nearly coincide at $r_s \sim 80$ where the initial wave is nonzero, these slices differ considerably in the strong field region near the black hole. If we wish to compare the results of the AHBC and perturbation methods in this region, we must use coordinate invariant quantities. We therefore introduce a set of stationary observers at specified values of the areal radius $r_s$, and we record the values of $\phi$ and $\phi_{,\tau}$ measured by each of these observers as a function of $\tau$, the proper time measured by the clock of each observer. To synchronize the clocks in a manner independent of the choice of time slicing, we introduce an ingoing light signal that passes $r_s = 80$ at $t_s = t = 0$. Each observer starts his clock running from $\tau = 0$ when he sees the signal, *e.g.*, at the initial time slice, an observer at $r_s = 80$ reads $\tau = 0$, an observer at $r_s = 90$ reads $\tau \sim 10$ (plus $O(1/r_s)$ corrections), and an observer at $r_s = 70$ reads $\tau \sim -10$.

Figure 6 shows $\phi$ and $\phi_{,\tau}$ as measured by two different observers. The observer at $r_s = 100M$ sees two peaks of scalar radiation. The first travels outward from $r_s = 80M$, starting at the observer's proper time $\tau \sim 20M$, and passes the observer at $\tau \sim 40M$. The second travels inward from $r_s = 80M$, is partially reflected by the black hole, and then moves outward, passing the observer at $\tau \sim 220M$. The observer at $r_s = 5M$ does not see the outgoing radiation pulse, but instead sees a combination of the ingoing pulse and its reflection. The agreement between the two numerical methods is excellent, demonstrating that the AHBC scheme is able to handle gravitational radiation without producing large numerical reflections at the apparent horizon or at the outer grid boundary. In addition, the above case extends from $t = 0$ to $t = 300M$, much longer than a traditional SA scheme would allow.

Figure 7 shows the coordinate velocity of outgoing light rays,

$$\frac{dr}{dt} = \frac{\alpha}{A} - \beta. \tag{5.6}$$

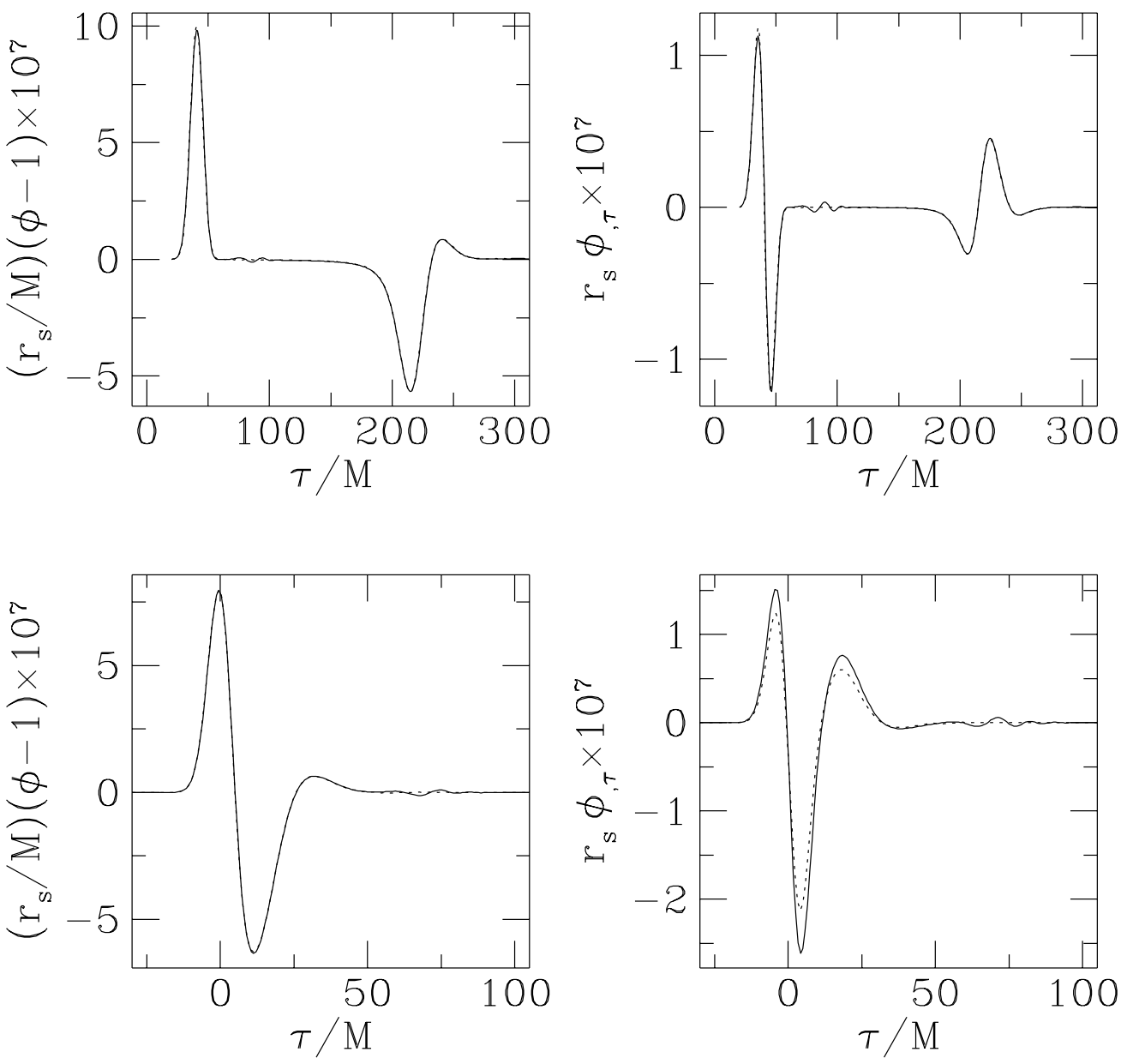

**Figure 6** Scalar field and its derivative with respect to proper time $\tau$, as measured by stationary observers, for a small scalar perturbation about a Schwarzschild black hole. The proper time $\tau$ is measured in units of $M$, the mass of the black hole. The AHBC method (solid line) and the perturbation method (dotted line) almost coincide. Both methods use 256 grid points. The observation radii, from top to bottom, are $r_s = 100M$ and $5M$.

This quantity is negative inside the apparent horizon, indicating that in this region information can only move inward with respect to the coordinates. Such a coordinate system is essential for solving the wave equation without explicitly imposing a boundary condition at the inner edge of the numerical grid. Because the spacetime is Schwarzschild, the apparent horizon coincides with the event horizon, so that the coordinate grid point on the AH moves along an outgoing light ray. At large radii $dr/dt$ approaches unity because the coordinate system is asymptotically Minkowskian.

## C. Oscillating Einstein Cluster

In order to demonstrate the emission of monopole gravitational radiation, we evolve a system that undergoes spherical oscillations. At $t = 0$ we place particles in randomly oriented stable circular orbits, and then reduce each particle's four-velocity component $u_\theta$ by a constant factor $\xi$. The particles are arranged so that the comoving energy density is initially uniform throughout the configuration. If $\xi = 1$, the particle distribution would remain in equilibrium. In general relativity such an equilibrium system is called an Einstein[28] cluster. For $\xi < 1$ and a weak gravitational field, particles do not remain in equilibrium, but instead move on elliptical orbits with identical periods, and the entire spherical distribution periodically expands and contracts in a homologous manner[29]. In strong gravitational fields, however, the oscillations of the distribution are not homologous because of shell crossing. In addition, the existence of a scalar field in Brans-Dicke theory allows monopole radiation, so that the system loses energy. The particles eventually settle into a new equilibrium state.

Figure 8 shows $\phi - 1$, $\Pi$, and the metric quantities $A^{1/2} - 1$ and $\alpha - 1$, measured at $r = 400M_0$ as a function of coordinate time for an oscillating cluster with $\xi = 0.95$ and an initial areal radius $R = 10M_0$. Here $M_0$ is the initial active gravitational mass $M$ of the cluster. The particles and the spacetime are evolved using the SA method. In order to produce a significant amount of gravitational radiation, we set $\omega = 1$. As the cluster oscillates, it produces scalar gravitational radiation that first reaches $r = 400M_0$ at a time $t \sim 400M_0$. The amplitude of the oscillation is damped because energy is lost to radiation. Notice that the lapse function $\alpha$ oscillates even before the radiation reaches $r = 400M_0$. This is because $\alpha$ is determined by elliptic equations which "feel" the oscillations of the interior matter distribution as well as the gravitational radiation that propagates to infinity.

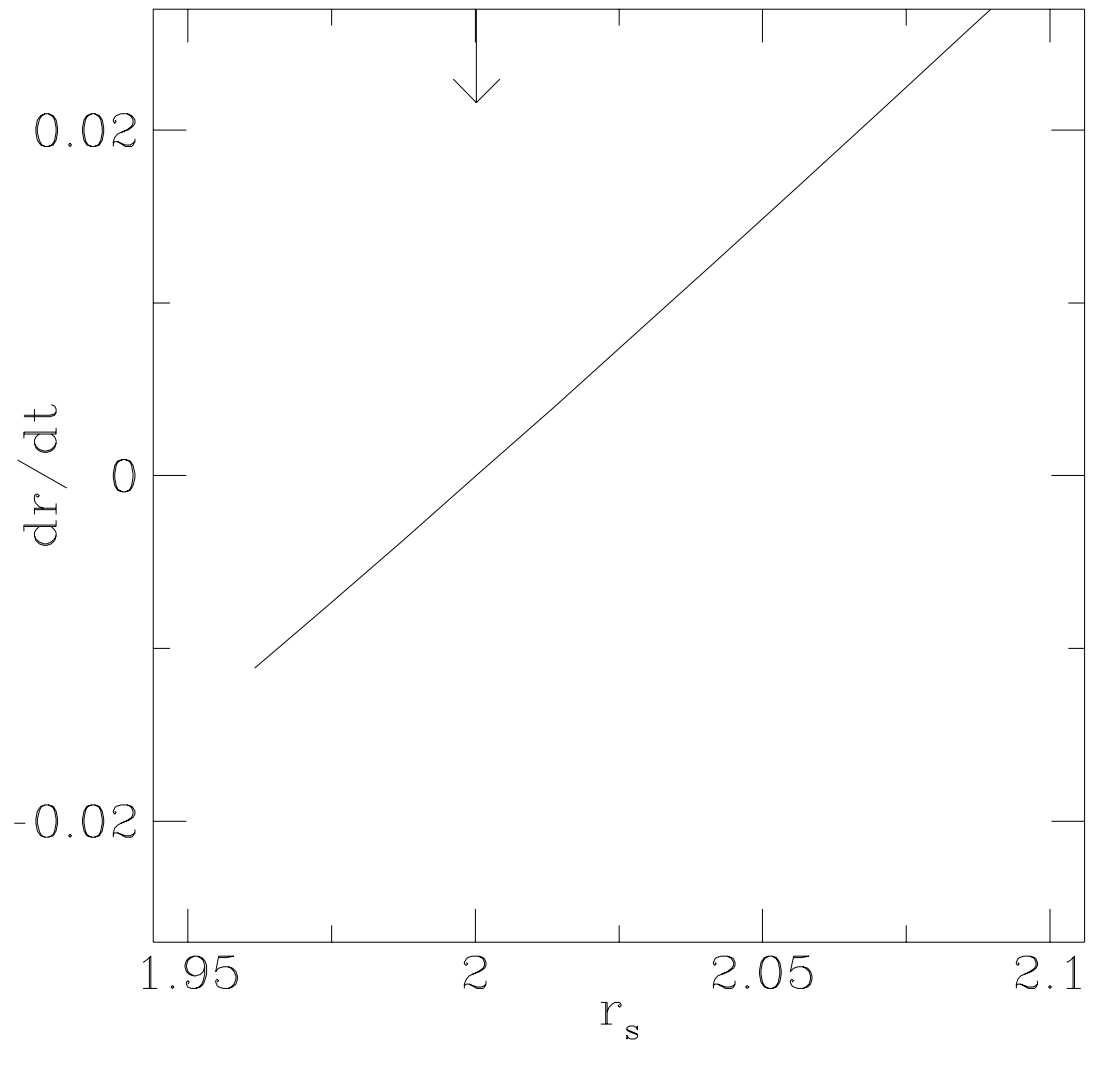

**Figure 7** Coordinate velocity $dr/dt$ of outgoing light rays versus areal radius $r_s$ for a Schwarzschild black hole with a weak scalar perturbation. Areal radius is measured in units of $M$, the mass of the black hole. The quantity $dr/dt$ is time-independent because the scalar field is too small to significantly change the background Schwarzschild metric. The arrow indicates the position of the apparent horizon, which is at $r_s = 2M$. The inner edge of the computational domain is at $r_s = 1.96M$.

### D. Oppenheimer-Snyder Collapse in Brans-Dicke Theory

We now consider a case in which the effects of the scalar field are large enough that one cannot use linear approximations to determine the evolution of the spacetime. We treat the analogue of Oppenheimer-Snyder collapse in Brans-Dicke theory. This case is identical to the one in section A except that we set $\omega = 1$ to increase the effect of the scalar field on the spacetime geometry. This value of $\omega$ conflicts with experiment; however, our purpose here is not to calculate a physically realistic collapse, but to provide a test of our numerical method.

We begin with a uniform, momentarily static sphere of noninteracting particles with total mass $M_0$ and initial radius $r_s = 10M_0$. The zero subscript on the mass refers to the initial value—as it evolves in time, the system loses mass because of scalar gravitational radiation. Because $\omega = 1$, the scalar mass $M_S$ contributes about 16% of the initial mass $M_0$. We follow the collapse of the configuration with our SA method, using 81 interior grid points, 175 exterior grid points, and 1200 particles. Our outer grid boundary is at $r = 100M_0$.

A spacetime diagram of the collapse is shown in Figure 9. The apparent horizon forms outside the matter surface at $t = 44M_0$. It initially has a radius $r_s = 1.42M_0$, but it expands as the scalar field radiates away. By $t = 65M_0$ the AH has grown to $r_s = 1.69M_0$. It then decreases by about $0.01M_0$ (not visible in the figure) and reaches equilibrium. This decrease represents a feature of Brans-Dicke black holes that is not present in general relativity. It is discussed further in Paper II. After $t = 90M_0$, numerical errors from exponentially growing metric gradients become large enough that the matter and AH both move outward in areal radius. These errors eventually cause the code to crash at $t = 99M_0$. Adding more grid points does not allow the code to run much longer: with twice as many grid points, it only runs until $t = 108M_0$.

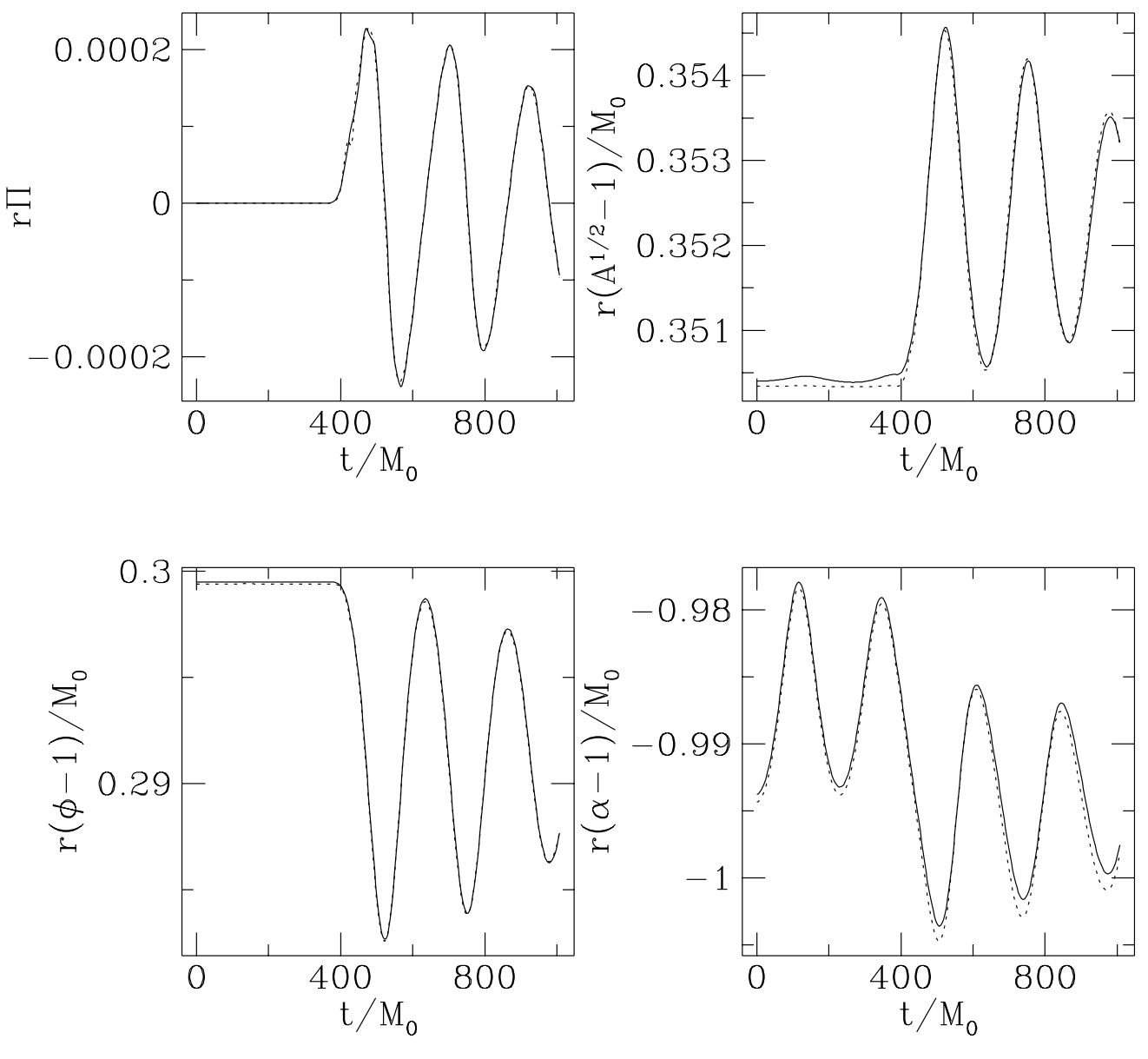

**Figure 8** Scalar field and metric quantities at $r = 400M_0$ for an oscillating Einstein cluster with $\omega = 1$, $\xi = 0.95$, and an initial areal radius $R = 10M_0$. The solid lines show results calculated with an outer grid radius at $r = 410M_0$, 1200 particles, 61 inner grid points, and 195 outer grid points. The dotted lines show results calculated with an outer grid radius at $r = 1000M_0$, 1200 particles, 61 inner grid points, and 451 outer grid points. The close agreement between the two cases demonstrates that our code accurately handles the outgoing wave boundary conditions at our outermost grid point, even while waves are passing through the boundary.

Figure 10 shows the scalar field as a function of areal radius and coordinate time. The scalar field begins radiating away soon after the formation of the apparent horizon at $t = 44M_0$. Outside the AH, a pulse of scalar radiation propagates outward towards infinity. Inside the AH, the value of $\phi$ is frozen in time because of the "collapse of the lapse". A large gradient in $\phi$ develops near the matter surface just outside $r_s/M = 1$. Until very late times, this region is well-covered by our grid, which follows the particles at every time step. However, the grid coverage in the exterior is extremely sparse: At $t = 80M_0$, only 18 of our 256 grid points lie outside coordinate radius $r/M_0 = 10$, and at $t = 90M_0$, only 29 of these grid points lie outside $r/M_0 = 1$. This makes it difficult to resolve the outgoing pulse of scalar radiation at late times.

Unlike the general relativistic case discussed in section V.A, the system has not settled into its final state by the time the SA code loses accuracy. This is because the spacetime remains dynamical after the formation of a black hole—it continues to emit scalar gravitational radiation. In addition, at $t = 80M_0$ one cannot accurately determine how much energy has been lost owing to gravitational radiation because the tail end of the waveform has not yet propagated far enough from the black hole to justify using linearized theory. By $t = 90M_0$, the waveform has propagated a bit further into the weak-field region, but the grid coverage in the exterior is becoming sparse.

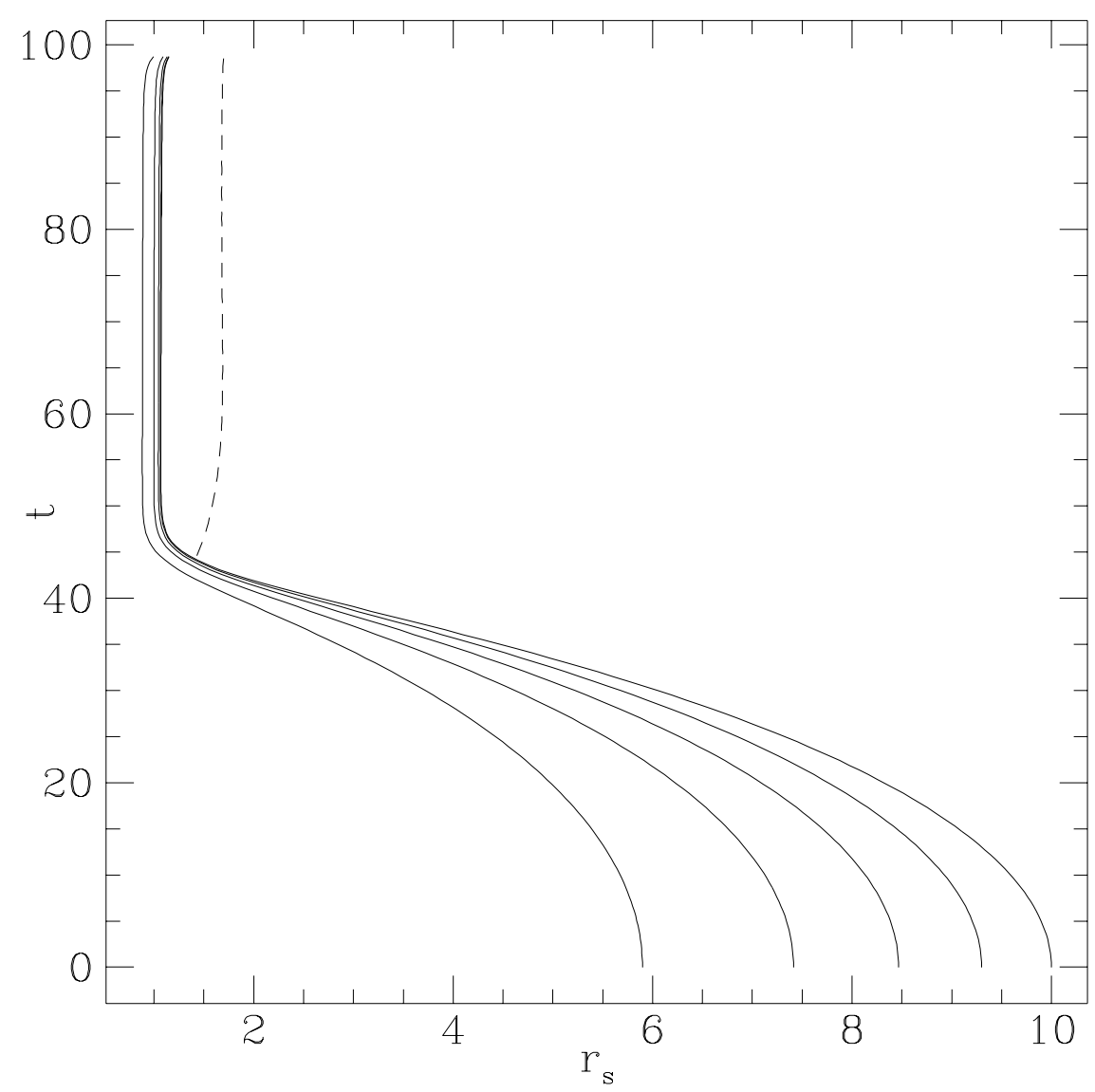

**Figure 9** Oppenheimer-Snyder Collapse in Brans-Dicke theory using maximal slicing, and calculated by the SA method. The five solid lines represent world lines of matter elements containing 20%, 40%, 60%, 80%, and 100% of the interior rest mass. Time $t$ and areal radius $r_s$ are measured in units of $M_0$, the initial mass of the configuration. The dotted line is the apparent horizon, which forms at about $t = 44M_0$.

The SA method fails here because it cannot maintain accuracy far enough into the future. Although one might be able to determine the waveform and radiated energy by increasing the grid coverage by a factor of 4 or so, this would be extremely costly in terms of computer time. The above simulation with 256 grid points requires 8 hours of CPU time on a SUN Sparcstation 5 to integrate to $t = 90M_0$. With 512 points, it takes 32 CPU hours to get to $t = 90M_0$, and 6 additional CPU hours to reach $t = 100M_0$. At each time step, the computer spends most of its time moving the particles and calculating source terms. The length of each time step is small ($\Delta t \sim 2 \times 10^{-3} M_0$ at late times, using 256 grid points) because of the Courant limit.

In the AHBC method we no longer have an exponentially varying metric, so we are able to integrate much farther into the future. The savings in computer time are enormous: Because we no longer have a Courant limitation and we no longer need to move particles, the AHBC method with 256 grid points requires only 20 minutes of CPU time to run from $t = 45M_0$ to $t = 300M_0$. The average length of each timestep is $\Delta t \sim 0.09M_0$, a factor of 45 larger than in the SA method for the same number of grid points. Furthermore, each step in the AHBC method takes an average of 0.4 CPU seconds, as opposed to about 1 CPU second for the SA method.

In Figure 11 we show the evolution of $\phi$ calculated using the AHBC method with 256 grid points. The initial data is provided by an SA slice at $t = 45M_0$. We show results only up to $t = 216M_0$, but our code is capable of running indefinitely. One can see the scalar field radiate away to infinity after the black hole forms. At the end of the simulation, we are left with the Schwarzschild solution and a constant scalar field, in agreement with Hawking's theorem[30] and with the simulations[4] of Shibata *et al.*.

Mass conservation is shown in Figure 12. The upper two plots show $M_{\rm T}$ and $M_{\rm S}$ at different radii, calculated from Eqs. (2.64) and (2.66). Notice that values of the tensor mass measured at different radii do not agree, even at the initial time step. The same is true for the scalar mass, although it is difficult to see this from the figure because the scale of the $M_{\rm S}$ plot is much larger than that of the $M_{\rm T}$ plot. This small discrepancy is not due to numerical error; it results from second-order terms that are not accounted for in

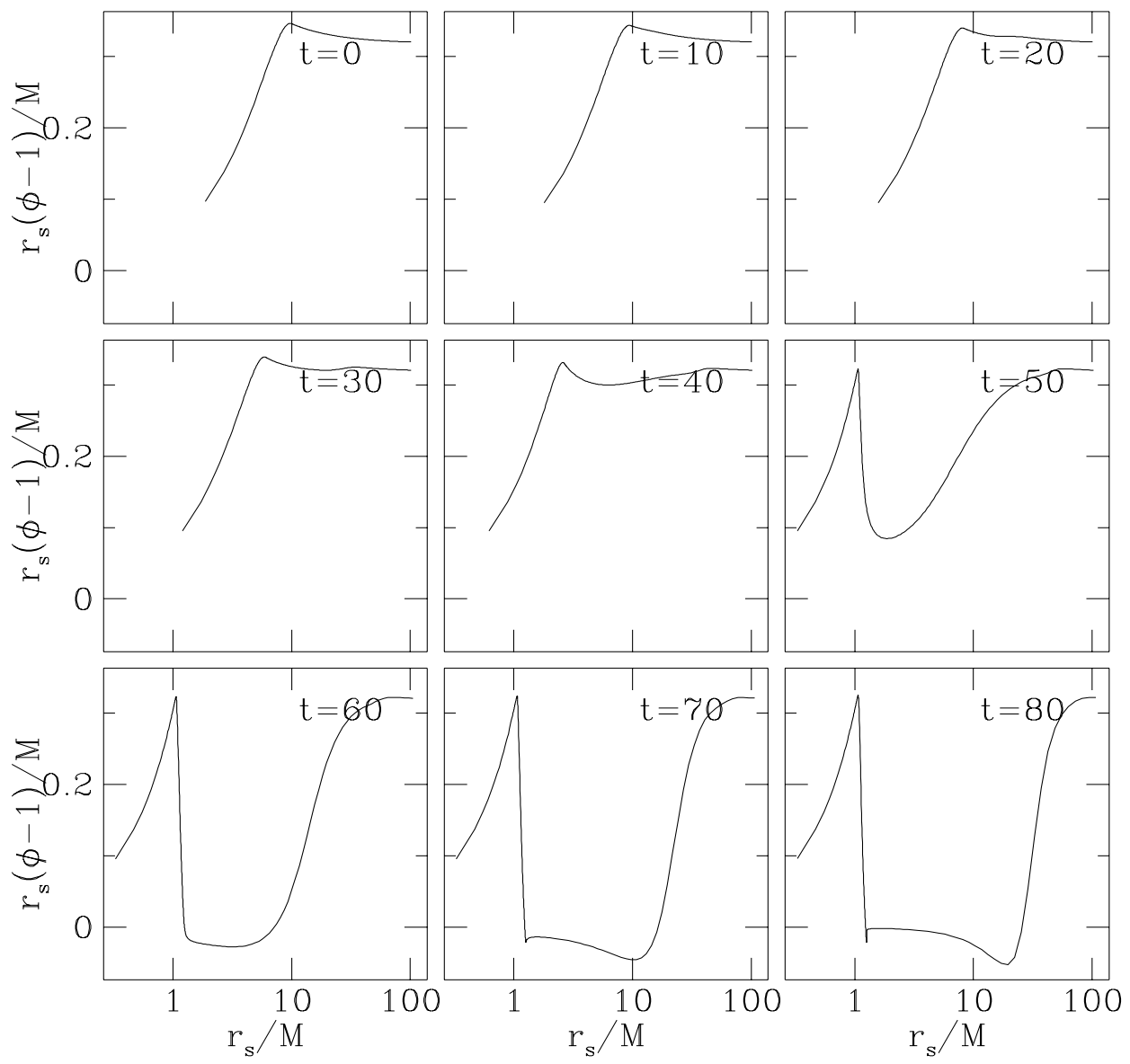

**Figure 10** Scalar field $\phi$ as a function of areal radius $r_s$ on selected maximal time slices for the collapse shown in Figure 9. Time and distance are measured in units of $M_0$. We plot $r_s/M_0$ on a logarithmic scale in order to show the region inside the AH.

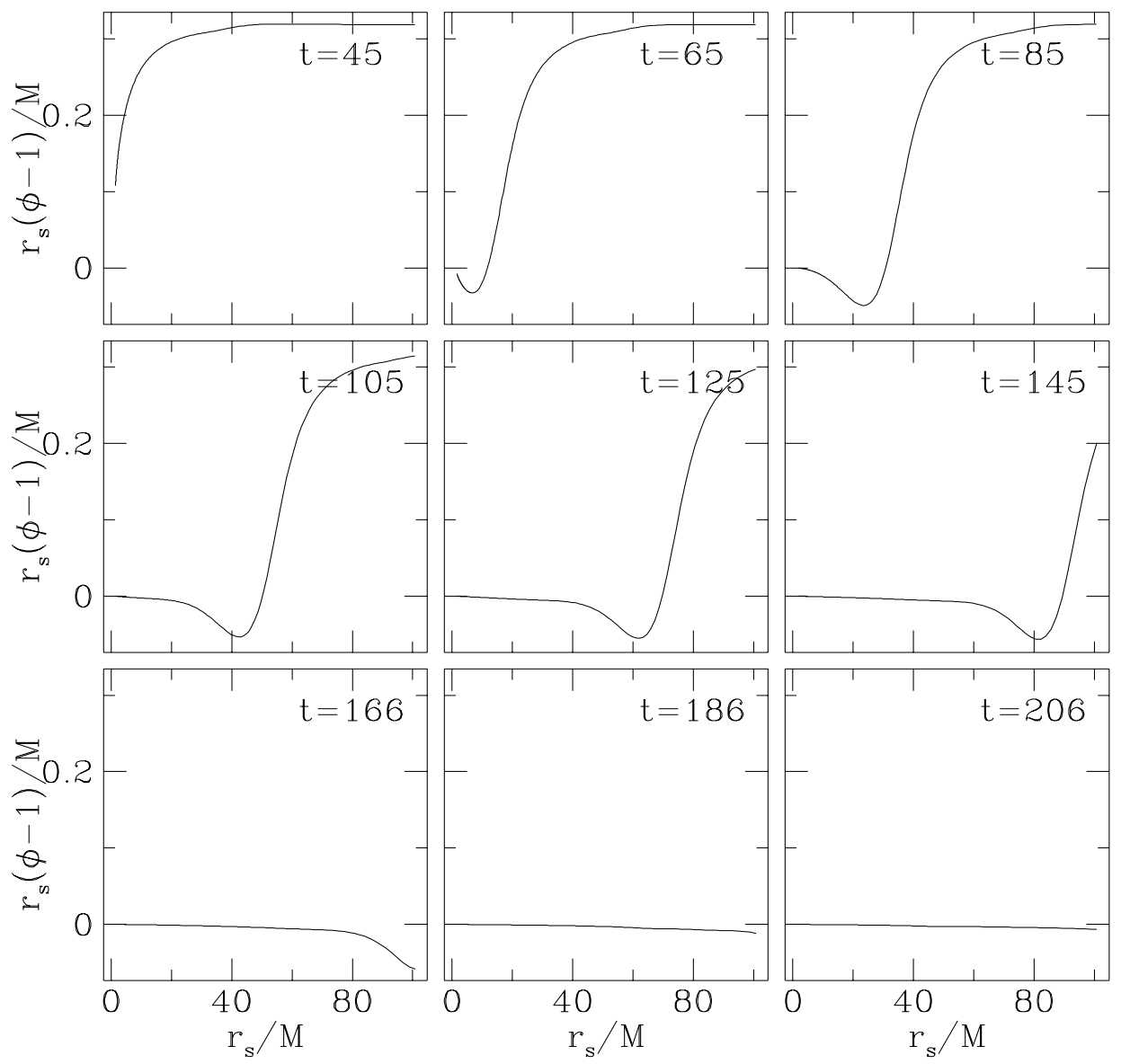

**Figure 11** Same as Figure 10, except calculated using the AHBC scheme, which allows one to integrate farther into the future. Time and distance are measured in units of $M_0$. The areal radius $r_s$ is plotted on a linear scale.

the linear-order expressions (2.64). These terms produce $O(1/r)$ corrections to the masses. It is interesting to note that the instantaneous values of $M_{\rm T}$ do not exhibit this error after the scalar field has radiated

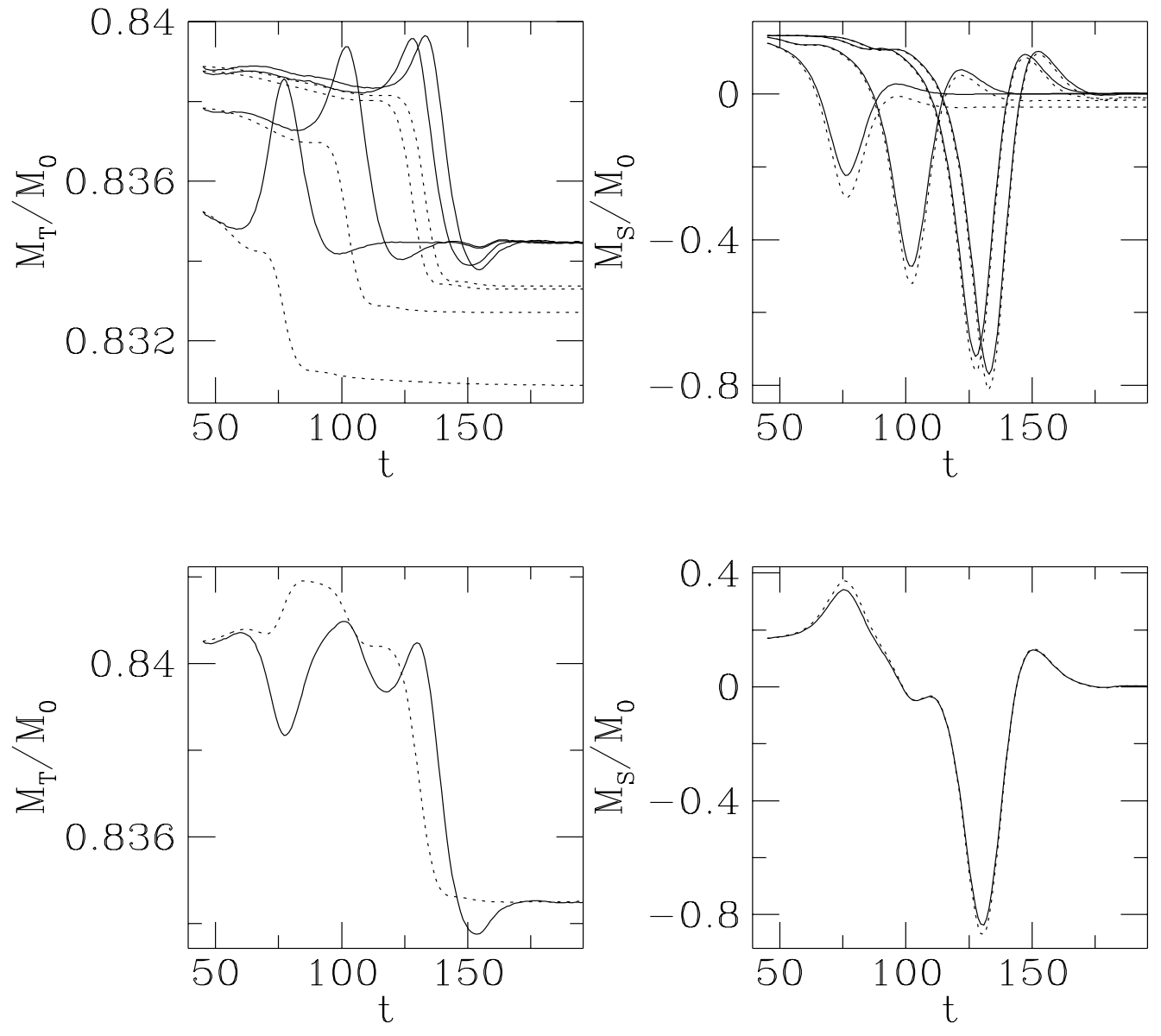

**Figure 12** Mass conservation for the AHBC method. In the upper two plots, the instantaneous masses from Eqs. (2.64) (solid lines) and the time-integrated masses from Eqs. (2.66) (dotted lines) are shown versus coordinate time at coordinate radii $80M_0$, $75M_0$, $50M_0$, and $25M_0$. Observers at smaller radii see the outgoing pulse at earlier times. The lower two plots are obtained by extrapolating the data in the upper two plots to $r = \infty$. Time is measured in units of $M_0$.

away—the solid lines in the upper left plot converge to the same value for $t > 175M_0$. This is because in the final state the spacetime is Schwarzschild, and in this case the second-order terms vanish[18]:

$$A^{1/2} = 1 + \frac{M}{2r} + O\left(\frac{1}{r^4}\right), \qquad \text{GR, vacuum.} \tag{5.7}$$

To remove the effects of the second-order terms and to determine the masses that one would measure at infinity, we extrapolate radially to $r = \infty$ by fitting to the form $M = C_1 + C_2/r$. The result is shown in the lower plots of Figure 12. Because this is a radial extrapolation with no time dependence, it is only meaningful in the initial and final states. In the final state, the instantaneous and time-integrated values agree remarkably well for both $M_T$ and $M_S$. This is a nontrivial result that verifies the accuracy of our numerical code.

The final mass of the configuration is $M = M_T + M_S \sim 0.83M_0$—the system loses about 17% of its initial mass because of scalar gravitational radiation. While all of the scalar mass is radiated away during the collapse, the loss of tensor mass is small, accounting for only about 3% of the mass loss. Although the tensor mass should decrease monotonically with the emission of gravitational radiation[17], the instantaneous values in Figure 12 do not. This is another effect of the second-order terms that are not included in the linearized equations.

## CONCLUSION

By adopting an AHBC method we have avoided the troublesome coordinate pathologies of singularity-avoiding schemes used in numerical relativity. We have used this method to evolve spherically symmetric spacetimes in Brans-Dicke theory. In particular, we have treated Oppenheimer-Snyder collapse to a black hole in both the general relativistic limit and in a regime in which the scalar field effects are strong. We have also tested our code against an independent method by evolving scalar perturbations on a Schwarzschild background. Our code can handle black hole spacetimes that contain gravitational radiation and is capable of maintaining high accuracy for an arbitrarily long time. Although we have restricted ourselves to spherical spacetimes, our scheme does not explicitly utilize results specific to spherical symmetry (such as matching to an exterior Schwarzschild metric). Accordingly, we feel that generalizing the scheme to multidimensions holds considerable promise.

## ACKNOWLEDGMENTS

We would like to thank A. Abrahams and T. Baumgarte for helpful discussions. This work has been supported in part by National Science Foundation grants AST 91-19475 and PHY 90-07834 and the Grand Challenge grant NSF PHY 93-18152/ ASC 93-18152 (ARPA supplemented).

## APPENDIX A. FINITE DIFFERENCE EQUATIONS FOR SA METHOD

In this appendix we discuss the numerical details of the SA method, including the finite difference approximations for equations found in section II. The numerical grid extends from $r_1 = 0$ to $r_{i_{\max}} = r_{\max}$. All variables are centered at half-grid points $r_{i+1/2}$. Time steps are labeled by the index $n$. We define the time derivative operator

$$\begin{aligned}\mathcal{T}[Y]^n_{i+1/2} &\equiv (Y_{,t})^n_{i+1/2} \\ &\equiv \frac{\Delta t_{n-1}}{\Delta t_n + \Delta t_{n-1}} \frac{Y^{n+1}_{i+1/2} - Y^n_{i+1/2}}{\Delta t_n} + \frac{\Delta t_n}{\Delta t_n + \Delta t_{n-1}} \frac{Y^n_{i+1/2} - Y^{n-1}_{i+1/2}}{\Delta t_{n-1}},\end{aligned} \tag{A1}$$

and a Laplacian-like operator that takes two arguments:

$$\begin{aligned}\mathcal{R}[X,Y]^n_{i+1/2} &\equiv \left(\left(XY_{,r^2}\right)_{,r^3}\right)^n_{i+1/2} \\ &\equiv \frac{1}{r^3_{i+1} - r^3_i}\left[X^n_{i+1}\left(\frac{Y^n_{i+3/2} - Y^n_{i+1/2}}{r^2_{i+3/2} - r^2_{i+1/2}}\right) - X^n_i\left(\frac{Y^n_{i+1/2} - Y^n_{i-1/2}}{r^2_{i+1/2} - r^2_{i-1/2}}\right)\right].\end{aligned} \tag{A2}$$

Here $\Delta t_n \equiv t_{n+1} - t_n$, and $X$ and $Y$ are any variables defined at half-grid points $r_{i+1/2}$. The quantities $X^n_i$ that appear in the finite difference expression for $\mathcal{R}$ are obtained from $X^n_{i+1/2}$ by linear interpolation in $r^2$.

We also define an operator for a derivative with respect to $r^2$:

$$\mathcal{Q}[Y]^n_{i+1/2} \equiv \left(Y_{,r^2}\right)^n_{i+1/2}. \tag{A3}$$

This operator will only be used for variables that satisfy

$$Y_{,r^2} \sim \text{constant}_1 + \text{constant}_2 \cdot r^2. \tag{A4}$$

Because all quantities are defined at half-grid points, the finite difference approximation of $\mathcal{Q}[Y]^n_{i+1/2}$ is obtained by first calculating

$$\left(Y_{,r^2}\right)^n_i \equiv \frac{Y^n_{i+1/2} - Y^n_{i-1/2}}{r^2_{i+1/2} - r^2_{i-1/2}} \tag{A5}$$

at all grid points $r_i$ and then interpolating linearly in $r^2$ to the half-grid points $r_{i+1/2}$. To determine $\mathcal{Q}[Y]^n_{i+1/2}$ at $i = 1$ we extrapolate linearly in $r^2$ using the values of $\left(Y_{,r^2}\right)^n_i$ at $i = 2$ and $i = 3$.

### 1. Wave Equation

The finite difference form of Eqs. (3.14) can be written

$$\mathcal{T}[\xi]^n_{i+1/2} = 2r_{i+1/2}\beta^n_{i+1/2}\mathcal{Q}[\xi]^n_{i+1/2} - \alpha^n_{i+1/2}\Pi^n_{i+1/2}, \tag{A6a}$$

$$\mathcal{T}[\Pi]^n_{i+1/2} = 2r_{i+1/2}\beta^n_{i+1/2}\mathcal{Q}[\Pi]^n_{i+1/2} - \frac{6}{(A^n_{i+1/2})^3}\mathcal{R}\left[r^3 A\alpha, \xi\right]^n_{i+1/2} + \frac{8\pi\tilde{T}^n_{i+1/2}\alpha^n_{i+1/2}}{(A^n_{i+1/2})^3(3+2\omega)}. \tag{A6b}$$

Notice that all quantities on the right-hand side of Eqs. (A6) are evaluated at timestep $n$. For a grid uniform in $r$, these equations are second-order accurate in both space and time.

Because of the way we have finite differenced the equations, we have already included the regularity condition (3.15) at the origin, so this equation need not be imposed explicitly. A finite difference approximation of the boundary condition (3.17), which is imposed at the outermost grid point, is

$$Y^{n+1}_{i_{\max}+1/2} = \frac{r_{i_{\max}-1/2}}{r_{i_{\max}+1/2}}Y^n_{i_{\max}-1/2} + \left(\frac{1-\zeta}{1+\zeta}\right)\left(Y^n_{i_{\max}+1/2} - \frac{r_{i_{\max}-1/2}}{r_{i_{\max}+1/2}}Y^{n+1}_{i_{\max}-1/2}\right), \tag{A7}$$

where

$$\zeta \equiv \frac{\Delta t_n}{r_{i_{\max}+1/2} - r_{i_{\max}-1/2}}, \tag{A8}$$

and $Y$ is either $\Pi$ or $\xi$. This equation is second-order accurate in space and time.

For a uniform Cartesian grid, the Courant stability condition for the above difference scheme is

$$\Delta t < \frac{\Delta r}{|\beta \pm 2\alpha/A|}, \tag{A9}$$

where $\Delta r \equiv r_{i+1} - r_i$. For a variable timestep and a nonuniform radial grid, we use the stability criterion

$$\Delta t = \varepsilon \min_i \left\{\frac{\Delta r_i}{|\beta_i \pm 2\alpha_i/A_i|}\right\}, \tag{A10}$$

where typically we choose $\varepsilon = 0.5$ for accuracy.

### 2. Constraints

The initial guess for $\phi$ is computed from Eq. (3.24). We use the finite difference scheme

$$\mathcal{T}[\psi]^n_{i+1/2} = 2r_{i+1/2}\beta^n_{i+1/2}\mathcal{Q}[\psi]^n_{i+1/2} + \frac{\beta^n_{i+1/2}\psi^n_{i+1/2}}{2r_{i+1/2}} - \frac{1}{4}\alpha^n_{i+1/2}\psi^n_{i+1/2}(K_{\mathrm{T}})^n_{i+1/2}. \tag{A11}$$

Since the evolution equation is only used for an initial guess, the stability of the scheme used to solve it is irrelevant, and the accuracy is only important for reducing the number of iterations needed to solve the constraints.

A second-order accurate finite-difference form of the momentum constraint (3.20) is

$$\frac{Z_{i+1/2} - Z_{i-1/2}}{r^5_{i+1/2} - r^5_{i-1/2}} = \frac{8\pi(\tilde{S}_r)_i}{5r_i} - \frac{2}{5}\psi_i^6\frac{\Pi_{i+1/2} - \Pi_{i-1/2}}{r^2_{i+1/2} - r^2_{i-1/2}} - \left(\frac{2\omega\psi_i^6}{5\phi_i}\right)\frac{\xi_{i+1/2} - \xi_{i-1/2}}{r^2_{i+1/2} - r^2_{i-1/2}}. \tag{A12}$$

The quantities $A_i$, $\Pi_i$, and $\phi_i$ defined on grid points $i$ are determined from their values on half-grid points using linear interpolation in $r^2$. To determine $Z_{3/2}$, we integrate from $r=0$ to $r=r_{3/2}$ using the fact that $\psi$, $\xi$, $\phi$, and $S_r/r$ obey Eq. (A4) near the origin. The result is

$$Z_{3/2} = \frac{\psi_{3/2}^6 r_{3/2}^4}{5}\left[8\pi (S_r)_{3/2} - 2r_{3/2}\left(\frac{\Pi_{5/2}-\Pi_{3/2}}{r_{5/2}^2 - r_{3/2}^2} + \left(\frac{\omega \Pi_{3/2}}{\phi_{3/2}}\right)\frac{\xi_{5/2}-\xi_{3/2}}{r_{5/2}^2 - r_{3/2}^2}\right)\right]. \tag{A13}$$

Eqs. (A12) and (A13) are solved for $Z$ by starting at $Z_{3/2}$ and working up to $Z_{i_{\max}+1/2}$. Because Eqs. (A12) and (A13) depend on $\psi$ through the scalar wave terms, we must solve this equation at every step in the iteration of the Hamiltonian constraint.

The Hamiltonian constraint, Eq. (3.21), can be written

$$\frac{6}{\phi^{1/2}}\left(r^3\phi^{1/2}\psi_{,r^2}\right)_{,r^3} = \sum_k R^{(k)}\psi^{p(k)}, \tag{A14}$$

where the coefficients $R^{(k)}$ do not depend on $\psi$. We solve this nonlinear equation by an iterative scheme. Let $\hat{\psi}$ be an initial guess for $\psi$, and write $\psi = \hat{\psi}(1+(\psi-\hat{\psi})/\hat{\psi})$. If we have a good initial guess, we can evaluate the right-hand side of Eq. (A14) to first order in $(\psi-\hat{\psi})/\hat{\psi}$. The result is

$$6\left(r^3\phi^{1/2}\psi_{,r^2}\right)_{,r^3} = \phi^{1/2}\sum_k R^{(k)}\hat{\psi}^{p(k)-1}\left(p(k)\psi + (1-p(k))\hat{\psi}\right). \tag{A15}$$

We finite difference this equation as follows:

$$\begin{aligned} 6\,\mathcal{R}\left[r^3\phi^{1/2},\psi\right]_{i+1/2} \\ = \phi_{i+1/2}^{1/2}\sum_k R_{i+1/2}^{(k)}\hat{\psi}_{i+1/2}^{p(k)-1}\left(p(k)\psi_{i+1/2} + (1-p(k))\hat{\psi}_{i+1/2}\right). \end{aligned} \tag{A16}$$

This is a tridiagonal set of linear equations that can be solved for $\psi$. The coefficients $R_{i+1/2}^{(k)}$ appearing in Eq. (A16) are given by

$$R_{i+1/2}^{(1)} = -\frac{3Z_{i+1/2}^2}{16 r_{i+1/2}^6 \phi_{i+1/2}^2}, \tag{A17a}$$

$$R_{i+1/2}^{(2)} = -\frac{2\pi\tilde{\rho}_{i+1/2}}{\phi_{i+1/2}}, \tag{A17b}$$

$$R_{i+1/2}^{(3)} = -\frac{\omega\Pi_{i+1/2}^2}{8\phi_{i+1/2}^2}, \tag{A17c}$$

$$R_{i+1/2}^{(4)} = -\omega\left(\frac{r_{i+1/2}\mathcal{Q}[\xi]_{i+1/2}}{\phi_{i+1/2}}\right)^2, \tag{A17d}$$

$$R_{i+1/2}^{(5)} = -\frac{3}{2}\mathcal{R}\left[r^3,\xi\right]_{i+1/2}, \tag{A17e}$$

and the values of $p(k)$ for $k=1,\ldots,5$ are $-7$, $-1$, 5, 1, and 1, respectively.

The boundary condition at the origin, Eq. (3.22), is taken care of automatically by the finite difference scheme and does not need to be added explicitly. The boundary condition (3.23) is imposed at the outermost grid point. In finite difference form, this condition is

$$\frac{(r\psi)_{i_{\max}+1/2} - (r\psi)_{i_{\max}-1/2}}{r_{i_{\max}+1/2} - r_{i_{\max}-1/2}} = 1 - \frac{1}{8}\left((r\Pi)_{i_{\max}+1/2} - (r\Pi)_{i_{\max}-1/2}\right). \tag{A18}$$

After obtaining the initial guess from Eq. (A11), we iterate Eq. (A16) until convergence. Because the variable $\psi$ appears in Eq. (A12), we must recalculate $Z$ at every step in the iteration.

### 3. Lapse and Shift

Equation (3.25) is linear in $\alpha$, and can be solved using the finite difference approximation

$$\begin{aligned}&\frac{6}{(A_{i+1/2})^3}\mathcal{R}\left[r^3 A, \alpha\right]_{i+1/2} \\ &=\alpha_{i+1/2}\left(\frac{8\pi}{\phi_{i+1/2}}\left(\tilde{\rho}_{i+1/2}+\frac{\tilde{T}_{i+1/2}}{2+3/\omega}\right)+\frac{\omega\Pi^2_{i+1/2}}{\phi^2_{i+1/2}}+\frac{3}{2}(K_{\mathrm{T}}^2)_{i+1/2}\right).\end{aligned} \tag{A19}$$

This is a tridiagonal system of equations, which is easily solved by standard methods. The value of $A$ at a grid point $i$ is obtained by linear interpolation in $r^2$.

There is no need to explicitly impose a boundary condition at the origin. The boundary condition at infinity, Eq. (3.27), is imposed at the outermost grid point. A finite-difference version of this equation

$$\frac{(r\alpha)_{i_{\max}+1/2}-(r\alpha)_{i_{\max}-1/2}}{r_{i_{\max}+1/2}-r_{i_{\max}-1/2}}=1+\frac{1}{2}\left((r\Pi)_{i_{\max}+1/2}-(r\Pi)_{i_{\max}-1/2}\right). \tag{A20}$$

For the shift equation (2.24), we use the finite difference approximation

$$\frac{1}{r_{i+1/2}-r_{i-1/2}}\left(\frac{\beta_{i+1/2}}{r_{i+1/2}}-\frac{\beta_{i-1/2}}{r_{i-1/2}}\right)=-\frac{3}{2}\frac{\alpha_i}{r_i}(K_{\mathrm{T}})_i, \tag{A21}$$

where $\alpha$ and $K_{\mathrm{T}}$ at the grid points $r_i$ are calculated from their values at the half-grid points by linear interpolation in $r^2$. The finite difference form of the the boundary condition (3.28) is

$$\beta_{i_{\max}+1/2}=\frac{1}{2}\left(rK_{\mathrm{T}}\right)_{i_{\max}+1/2}+\frac{1}{2}\left(r\Pi\right)_{i_{\max}+1/2}. \tag{A22}$$

Eqs. (A21) and (A22) are easily solved by starting at $i=i_{\max}$ and proceeding down to $i=1$.

## APPENDIX B. INITIAL DATA FOR SA METHOD

If one chooses Schwarzschild coordinates such that

$$ds^2=-e^{2\Phi}dt^2+e^{2\Lambda}dr_s^2+r_s^2d\Omega^2, \tag{B1}$$

then the Brans-Dicke field equations for a *static* configuration of dust can be written in the form[23]

$$2\Lambda'=8\pi r_s e^{2\Lambda-\psi}\left(\rho+\frac{T}{3+2\omega}\right)+\frac{1+r_s\psi'}{2+r_s\psi'}\left(\frac{2-2e^{2\Lambda}}{r_s}\right)+\frac{(2+\omega)r_s\psi'^2}{2+r_s\psi'}, \tag{B2}$$

$$\psi''=-\frac{\psi'}{r_s}\left(1+e^{2\Lambda}\right)+8\pi e^{2\Lambda-\psi}\left[\frac{T}{3+2\omega}+\frac{r_s\psi'}{2}\left(\rho+\frac{T}{3+2\omega}\right)\right], \tag{B3}$$

$$\Phi'=\frac{1}{r_s(2+r_s\psi')}\left(-2r_s\psi'+\frac{1}{2}\left(r_s\psi'\right)^2+e^{\Lambda}-1\right), \tag{B4}$$

where

$$\psi \equiv \ln \phi. \tag{B5}$$

Here primes denote differentiation with respect to the areal radius $r_s$, and we have chosen units such that $\phi = 1$ at $r_s = \infty$.

For a static distribution of particles with comoving density $\rho_\star$, the source terms in the above equations are given by

$$\rho = \rho_\star (\tilde{u}^0)^2, \tag{B6}$$

$$T = -\rho_\star. \tag{B7}$$

Here

$$\tilde{u}^0 \equiv u^0 e^\Phi, \tag{B8}$$

where $u^0$ is the time component of the particles' mean four-velocity field. If all particles are static, $\tilde{u}^0 = 1$. If particles are in randomly oriented circular orbits,

$$\tilde{u}^0 = (1 - r_s \Phi')^{-1/2}. \tag{B9}$$

We can find the isotropic radius $r$ from the relation

$$r' = \frac{r e^\Lambda}{r_s}. \tag{B10}$$

The amount of rest mass $M_{\rm rest}$ enclosed within a radius $r_s$ is given by

$$M'_{\rm rest} = \frac{4\pi\rho e^\Lambda r_s^2}{\tilde{u}^0}. \tag{B11}$$

Near the center of the star, Eqs. (B2)–(B11) reduce to

$$\psi = \psi_0 + \frac{4\pi e^{-\psi_0} r_s^2}{3} \frac{T_0}{3 + 2\omega} + \cdots, \tag{B12}$$

$$\Lambda = \frac{4\pi e^{-\psi_0} r_s^2}{3} \left(\rho_0 + \frac{T_0}{3 + 2\omega}\right) + \cdots, \tag{B13}$$

$$\Phi = \Phi_0 + \frac{2\pi e^{-\psi_0} r_s^2}{3} \left(\rho_0 - \frac{T_0}{3 + 2\omega}\right) + \cdots, \tag{B14}$$

$$r = D r_s \left[1 + \frac{2\pi e^{-\psi_0} r_s^2}{3} \left(\rho_0 + \frac{T_0}{3 + 2\omega}\right) + \cdots\right], \tag{B15}$$

$$M_{\rm rest} = \frac{4\pi\rho_0 r_s^3}{3} + \cdots \qquad \text{(B16)},$$

where a zero subscript indicates a value at the center. The constants $D$, $\Phi_0$, and $\psi_0$ are determined by matching to the exterior solution.

For the exterior solution, we match to the Brans type I metric[31]. This solution can be expressed analytically in Schwarzschild-like coordinates[32]:

$$\phi = \xi^\chi, \tag{B17}$$

$$r_s = \frac{4B\xi^{1-Q}}{1 - \xi^2}, \tag{B18}$$

$$e^\Phi = \xi^{Q-\chi}, \tag{B19}$$

$$e^{-\Lambda} = \frac{1 + \xi^2 + Q\left(\xi^2 - 1\right)}{2\xi}, \tag{B20}$$

The constants $B$, $Q$, and $\chi$ are determined by matching to the interior solution, subject to the constraint

$$Q^2 + \chi^2 \left(1 + \frac{\omega}{2}\right) - \chi Q - 1 = 0. \tag{B21}$$

In the exterior, the isotropic radius is given by

$$r = B\frac{1+\xi}{1-\xi}. \tag{B22}$$

Note that our parameter $\chi$ corresponds to Matsuda's $A/\lambda$ and Brans' $C/\lambda$, and our parameter $Q$ corresponds to Matsuda's $(1+A)/\lambda$ and Brans' $(1+C)/\lambda$.

To determine the complete solution for a star with a given areal radius $R_s$ and a given density profile, we first integrate Eqs. (B2)–(B4) and (B10)–(B11) from $r_s = 0$ to $r_s = R_s$. These equations are ODEs and can be integrated to arbitrary accuracy by standard numerical methods. To perform the integration, we make an arbitrary choice for $\rho_\star(r_s = 0)$, and we set $D = 1$ and $\Phi_0 = \psi_0 = 0$. We will later determine the true values of $D$ and $\psi_0$ by matching to the exterior solution. The value of $\Phi_0$ is completely arbitrary, and only serves to determine the time coordinate.

After performing the integration, we match $e^\Lambda$ and $\psi'$ to the exterior solution, thereby obtaining $\xi_S$, the surface value of $\xi$, as well as the parameters $Q$ and $\chi$. The quantity $\xi_S$ is given by

$$a_1\left(\xi_S^2 + 1\right) + a_2\xi_S = 0, \tag{B23}$$

where

$$a_1 \equiv e^{-\Lambda_S}\left(1 + \frac{1}{2}r_s\psi_S'\right), \tag{B24}$$

$$a_2 \equiv -1 - e^{-2\Lambda_S}\left(1 + r_s\psi_S' + \left(1+\tfrac{1}{2}\omega\right)\left(r_s\psi_S'\right)^2\right). \tag{B25}$$

The constants $Q$ and $\chi$ are found from

$$Q = \frac{1 + \xi_S^2 - 2\xi e^{-\Lambda_S}}{1-\xi_S^2}, \tag{B26}$$

$$\chi = \frac{2r_s\psi_S'\xi_S e^{-\Lambda_S}}{1-\xi_S^2}. \tag{B27}$$

Knowledge of $\xi_S$ and $\chi$ determines the value of $\psi$ at the surface as calculated from the exterior solution:

$$\psi_{S_\mathrm{ext}} = \chi \ln \xi_S. \tag{B28}$$

In general, this will not be equal to the value of $\psi_S$ obtained from the interior solution, $\psi_{S_\mathrm{int}}$, because we chose $\psi(r_s = 0) = 0$ when integrating the ODE's. In order to match the value of $\psi$ at the surface, we define

$$\psi_0 \equiv \chi \ln \xi_S - \psi_{S_\mathrm{int}}, \tag{B29}$$

and make the transformation

$$\begin{aligned} \psi &\rightarrow \psi + \psi_0, \\ \rho &\rightarrow \rho e^{\psi_0}, \\ T &\rightarrow T e^{\psi_0}, \\ M_\mathrm{rest} &\rightarrow M_\mathrm{rest} e^{\psi_0}, \end{aligned} \tag{B30}$$

everywhere in the interior. We transform the variables $\rho$, $T$, and $M_{\rm rest}$ along with $\psi$ so that the quantities $\Lambda$, $\Phi$, $r$, $r_s$, and $\psi'$ remain invariant—this way we do not need to recalculate $\xi_S$, $Q$, and $\chi$ from Eqs. (B23)–(B27). This invariance is easily verified by examining Eqs. (B2)–(B4) and (B10)–(B11).

Finally, we determine the value of $B$ from Eq. (B18) evaluated at the surface

$$B = \frac{R_s(1-\xi_S^2)\xi_S^{Q-1}}{4}, \tag{B31}$$

and we obtain the value of $D$ by matching $r_s$ and $r$ at the surface:

$$D = \frac{B(1+\xi_S)}{R(1-\xi_S)}. \tag{B32}$$

Making the transformation $r \to Dr$ everywhere in the interior completes the solution.

The various masses are found from the exterior solution:

$$M_{\rm S} = -B\chi, \tag{B33}$$
$$M_{\rm T} = B(2Q-\chi), \tag{B34}$$
$$M = 2B(Q-\chi), \tag{B35}$$

Given an areal radius $R_s$, a density profile $\rho_\star(r_s)/\rho_\star(0)$, and a prescription for determining the four-velocities of particles such as Eq. (B9), the entire solution is a determined by only one parameter, the initial value of $\rho_\star(0)$ used to begin the integration (this is not the *true* value of $\rho_\star(0)$ because $\rho_\star(0)$ is modified by the transformation (B30)). To construct a solution with a particular mass, we vary the initial value of $\rho_\star(0)$ until Eq. (B35) yields the desired result. This procedure can be thought of as finding a root of a single (complicated) function of one variable.

## APPENDIX C. SOLUTION OF WAVE EQUATION IN AHBC METHOD

In the AHBC method, the coordinate velocity of an outgoing light ray, given by Eq. (2.68), is negative at coordinate grid points located inside the CCH. Therefore, in this region information propagates in the inward direction with respect to the coordinates. In principle, any quantity at a given grid point in this region is completely determined by information that has propagated inward from the exterior. This implies that no boundary condition should be imposed at the innermost edge of the grid, since this edge lies in the causal future of the remainder of the spacetime. We take advantage of this property by calculating quantities at a given grid point inside the CCH without using information from the region interior to that grid point. Not only does this permit us to use a numerical scheme in which information inside the CCH cannot move outward, but it also frees us from imposing an explicit boundary condition at the innermost grid radius.

Alcubierre[11] has constructed an implicit scheme for solving the scalar wave equation in flat 1+1 dimensional spacetime, but in an arbitrarily moving coordinate system. His scheme treates transitions between regions of the grid where light rays can move in only one coordinate direction, and regions in which they can move in both directions. In the former regions, his scheme ensures that information propagates in only one direction.

We use a method based on Alcubierre's analysis, but we do not use the same finite difference scheme. Like his method, ours is implicit, uses spatially averaged time derivatives, and requires no rewriting of the field equations in an inertial coordinate system. This makes it different from the causal differencing method of Seidel and Suen[9] and the causal reconnection scheme of Alcubierre and Schutz[10]. However, while Alcubierre solves the second order wave equation for $\phi$ using a scheme with three time levels, we solve two coupled first order equations (Eqs. 4.4 and 4.5) for the variables $\Phi$ and $\Pi$ using a two-level scheme. Directly solving second order equations (in time) is not very well suited for 3+1 numerical relativity, in which one prefers to have initial data defined on a single Cauchy surface, and one propagates this data from one time slice to the next.

### 1. Finite Difference Equations

We define the following time derivative operator $\mathcal{T}_{\mathcal{A}}$ that averages over the three nearest spatial grid points:

$$\mathcal{T}_{\mathcal{A}}[Y]_i^{n+1/2} \equiv (1-\theta_i(1+\lambda_i))\,\frac{\Pi_i^{n+1}-\Pi_i^n}{\Delta t} + \theta_i\,\frac{\Pi_{i+1}^{n+1}-\Pi_{i+1}^n}{\Delta t} + \theta_i\lambda_i\,\frac{\Pi_{i-1}^{n+1}-\Pi_{i-1}^n}{\Delta t}, \tag{C1}$$

where

$$\lambda_i \equiv \frac{\eta_{i+1}-\eta_i}{\eta_i-\eta_{i-1}}, \tag{C2}$$

$$\Delta t \equiv t^{n+1}-t^n. \tag{C3}$$

For a spatial grid uniform in $\eta$, we have $\lambda_i = 1$. The above operator is second order accurate in both space and time. The quantity $\theta_i$ is a numerical coefficient that describes the amount of spatial averaging, and is discussed further below. As discovered by Alcubierre, this averaging is important for solving the wave equation in the regime where the coordinate speed of outgoing light rays is negative. For $\theta_i = 0$ there is no averaging; the time derivative at spatial grid point $\eta_i$ is computed using only quantities at $\eta_i$.

In order to average quantities in time, we introduce the operator

$$\mathcal{A}_t[Y]_i^{n+1/2} \equiv \frac{1}{2}\left(Y_i^{n+1}+Y_i^n\right), \tag{C4}$$

which is accurate to second order in $\Delta t$. Because most of our variables, such as $\Pi$, are defined on the spatial grid points $\{i_1,\ldots,i_{\max}\}$, but the variable $\Phi$ is defined on the half grid points $\{i_{3/2},\ldots,i_{\max}+1/2\}$, we define an operator that averages over spatial grid points:

$$\mathcal{A}_\eta[Y]_{i+1/2}^n \equiv \frac{1}{2}\left(Y_{i+1}^n+Y_i^n\right). \tag{C5}$$

Because of our staggered grid, we use a three-point spatial derivative operator $\mathcal{D}$ and a two-point spatial derivative operator $\mathcal{D}_{1/2}$:

$$\mathcal{D}[Y]_i \equiv (Y_{,\eta})_i = \frac{\eta_i-\eta_{i-1}}{\eta_{i+1}-\eta_{i-1}}\left(\frac{Y_{i+1}-Y_i}{\eta_{i+1}-\eta_i}\right) + \frac{\eta_{i+1}-\eta_i}{\eta_{i+1}-\eta_{i-1}}\left(\frac{Y_i-Y_{i-1}}{\eta_i-\eta_{i-1}}\right), \tag{C6}$$

and

$$\mathcal{D}_{1/2}[Y]_i \equiv (Y_{,\eta})_i = \frac{Y_{i+1/2}-Y_{i-1/2}}{\eta_{i+1/2}-\eta_{i-1/2}}. \tag{C7}$$

The operator $\mathcal{D}$ is second order accurate even for a nonuniform spatial grid.

In vacuum, the wave equation (4.4)–(4.5) can be written in the following implicit finite difference form using the above operators:

$$\mathcal{T}_{\mathcal{A}}[\Pi]_i^{n+1/2} = Q\mathcal{D}[\mathcal{A}_t[\Pi]]_i^{n+1/2} + P\mathcal{D}_{1/2}[\mathcal{A}_t[\Phi]]_i^{n+1/2} + S\mathcal{A}_t[\mathcal{A}_\eta[\Phi]]_i^{n+1/2}, \tag{C8}$$

$$\mathcal{T}_{\mathcal{A}}[\Phi]_{i+1/2}^{n+1/2} = W\mathcal{D}[\mathcal{A}_t[\Phi]]_{i+1/2}^{n+1/2} + X\mathcal{A}_t[\Phi]_{i+1/2}^{n+1/2} + U\mathcal{D}_{1/2}[\mathcal{A}_t[\Pi]]_{i+1/2}^{n+1/2} + V\mathcal{A}_t[\mathcal{A}_\eta[\Pi]]_{i+1/2}^{n+1/2}. \tag{C9}$$

The coefficients $Q$, $P$, $S$, $W$, $T$, $U$, and $V$ are given by

$$Q = \frac{\beta_i^n}{r_i}, \tag{C10a}$$

$$P = -\frac{\alpha_i^n}{(\psi_i^n)^4 r_i}, \tag{C10b}$$

$$S = P\left(2 + \frac{1}{\alpha_i^n}\mathcal{D}[\alpha]_i^n + \frac{2}{\psi_i^n}\mathcal{D}[\psi]_i^n\right), \tag{C10c}$$

$$W = \frac{\mathcal{A}_\eta\,[\beta]_{i+1/2}^n}{r_{i+1/2}}, \tag{C10d}$$

$$X = \frac{1}{r_{i+1/2}}\mathcal{D}_{1/2}\,[\beta]_{i+1/2}^n, \tag{C10e}$$

$$U = -\frac{\mathcal{A}_\eta\,[\alpha]_{i+1/2}^n}{r_{i+1/2}}, \tag{C10f}$$

$$V = -\frac{1}{r_{i+1/2}}\mathcal{D}_{1/2}\,[\alpha]_{i+1/2}^n. \tag{C10g}$$

Notice that the coefficients defined by Eqs. (C10) are *not* centered at timestep $n+1/2$, but instead are centered at timestep $n$. This is because the quantities $\alpha$, $\beta$, and $\psi$ are only known at timestep $n$ when the wave equation is solved. As a result, this difference scheme is only first order accurate in time, although it remains second order accurate in space (for a uniform spatial grid). However, in the case where the metric coefficients change much more slowly than the wave variables $\Pi$ and $\Phi$, the scheme becomes second order accurate in both space and time.

If one neglects the terms with $S$, $X$, and $V$ in Eqs. (C8) and (C9), and one assumes a uniformly spaced grid, a Von Neumann analysis shows that the above difference scheme is unconditionally stable. The terms containing $S$, $X$, and $V$ should not significantly affect stability because these terms do not contain derivatives of $\Phi$ or $\Pi$.

The boundary conditions at the outer grid point, Eqs. (4.6) and (4.7), can be written

$$\frac{\Pi_i^{n+1} - \Pi_i^n}{\Delta t} = -\frac{1}{r_i}\mathcal{D}_{1/2}[\mathcal{A}_t[\Phi]]_i^{n+1/2} - \frac{2}{r_i}\mathcal{A}_\eta[\mathcal{A}_t[\Phi]]_i^{n+1/2}, \tag{C11}$$

$$\frac{\Phi_i^{n+1} - \Phi_i^n}{\Delta t} = -\frac{1}{r_i}\mathcal{D}_{1/2}[\mathcal{A}_t[\Phi]]_i^{n+1/2} - \frac{2}{r_i}\mathcal{A}_\eta[\mathcal{A}_t[\Phi]]_i^{n+1/2} + \frac{1}{r_i}\mathcal{A}_t[\Pi]. \tag{C12}$$

Both of these conditions are imposed at $i = i_{\max}$. For a grid uniform in $\eta$, they are second-order accurate in both space and time.

## 2. Causal Solution Method

To solve the wave equation at each time step, we first determine the location of the CCH to the nearest grid point using Eq. (2.68). We then define a causal boundary at $r_{i_{\rm CB}} = r_{\rm CB}$, which we place either at the CCH or the AH, whichever is smaller.

Eqs. (C8) and (C9) together with the boundary conditions (C11) and (C12) comprise a coupled system of linear equations for the variables $\Phi_{i+1/2}^{n+1}$ and $\Pi_i^{n+1}$, where $i = \{1, \ldots, i_{\max}\}$. Because there are a total of $2i_{\max}$ variables and only $2i_{\max} - 2$ equations, the system is underdetermined. However, this is a shortcoming of our finite difference approximation rather than a property of the underlying differential equations. In the continuum limit, Eqs. (4.4) and (4.5) together with the boundary conditions (4.6) and (4.7) should uniquely determine $\Phi$ and $\Pi$ everywhere in the spacetime region covered by our grid, given appropriate initial data. No additional boundary condition is needed at the inner edge of this region because no information can propagate outward from this boundary.

This leads us to the following questions: Which of the $2i_{\rm max}$ variables should be determined by the $2i_{\rm max} - 2$ finite difference equations, and how should we determine the remaining variables?

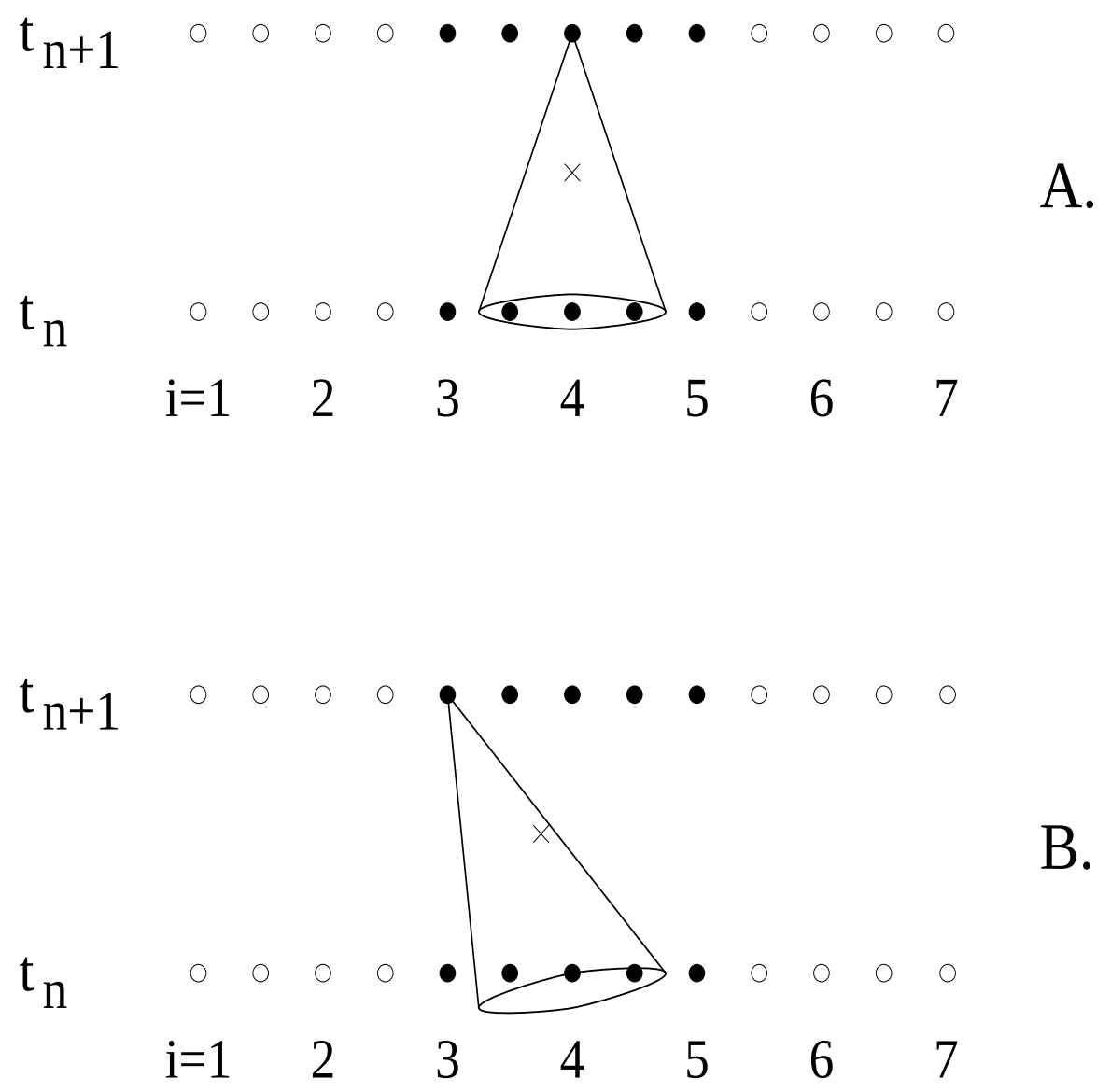

**Figure 13** Spacetime diagram showing the grid points involved in Eq. (C8) for $i = 4$. This equation is centered at the event $(i = 4, t = t_{n+1/2})$, indicated by a cross in the figure. The solid circles denote the grid points involved in the equation; other grid points are shown as open circles. In case A, left-directed and right-directed light rays move in opposite directions, as indicated by the light cone. In case B, left-directed and right-directed light rays move to the left with respect to the coordinates.

For the moment, consider only Eq. (C8). For a particular value of $i$, this equation involves 5 grid points at timestep $n$ and 5 grid points at timestep $n+1$. These are the points labeled by $i$, $i+1$, $i-1$, $i+1/2$, and $i-1/2$. Figure 13A is a spacetime diagram showing these grid points for the case $i = 4$, in a coordinate system where oppositely directed photons move in opposite directions on the grid. In this case, one traditionally uses Eq. (C8) to determine $\Pi_4^{n+1}$ in terms of quantities defined at the other nine grid points.

Now consider the case in Figure 13B, in which the coordinates are chosen so that both left-directed and right-directed light rays move to the left. We could proceed in the same way as we did in Figure 13A, and use Eq. (C8) for $i = 4$ to determine $\Pi_4^{n+1}$. This approach should cause no difficulty for either the stability or accuracy of the scheme. However, we instead choose to exploit the causal structure of the problem by using Eq. (C8) for $i = 4$ to determine $\Pi_3^{n+1}$. In this way, quantities at the point $(i = 3, t = t_{n+1})$ only depend on data from points with $i \geq 3$. This would not be permitted for the case shown in Figure 13A, since in that case, quantities at the point $(i = 3, t = t_{n+1})$ should be determined from information that propagates from both directions. However, in the case shown in Figure 13B, information can in principle only propagate to the left, a property which our scheme enforces.

We therefore adopt the following solution method: For grid points $i = 1, \ldots, i_{\rm CB}$, we solve for $\Pi_i^{n+1}$ and $\Phi_{i+1/2}^{n+1}$ using Eqs. (C8) and (C9) centered at $i+1$ and $i+3/2$, as in Figure 13B. For grid points $i = i_{\rm CB}+2, \ldots, i_{\max}-1$, we solve for $\Pi_i^{n+1}$ and $\Phi_{i+1/2}^{n+1}$ using Eqs. (C8) and (C9) centered at $i$ and $i+1/2$, as in Figure 13A. The quantities $\Pi_{i_{\max}}$ and $\Phi_{i_{\max}+1/2}$ are determined by the boundary conditions (C11) and (C12). Following Alcubierre, we determine the two remaining variables, $\Pi_{i_{\rm CB}+1}$ and $\Phi_{i_{\rm CB}+3/2}$, by requiring the functions $\Pi(r)$ and $\Phi(r)$ to be smooth: we use quadratic interpolation to obtain the two extra equations.

This procedure can be encoded into a single matrix equation, which we write schematically in the form

$$\left(\begin{array}{cccccccccccc}
\times & \bullet & \bullet & \bullet & \bullet & & & & & & & \\
 & \times & \bullet & \bullet & \bullet & \bullet & & & & & & \\
 & & & \ddots & & & & & & & & \\
 & & & \times & \bullet & \bullet & \bullet & \bullet & & & & \\
 & & & & \times & \bullet & \bullet & \bullet & \bullet & & & \\
 & & & \bullet & & \times & & \bullet & & \bullet & & \\
 & & & & \bullet & & \times & & \bullet & & \bullet & \\
 & & & & & \bullet & \bullet & \times & \bullet & \bullet & & \\
 & & & & & & \bullet & \bullet & \times & \bullet & \bullet & \\
 & & & & & & & & \ddots & & & \\
 & & & & & & & & & \bullet & \times & \bullet \\
 & & & & & & & & & \bullet & \bullet & \times
\end{array}\right)
\left(\begin{array}{c}
\Pi_1^{n+1} \\ \Phi_{3/2}^{n+1} \\ \vdots \\ \Pi_{i_{\rm CB}}^{n+1} \\ \Phi_{i_{\rm CB}+1/2}^{n+1} \\ \Pi_{i_{\rm CB}+1}^{n+1} \\ \Phi_{i_{\rm CB}+3/2}^{n+1} \\ \Pi_{i_{\rm CB}+2}^{n+1} \\ \Phi_{i_{\rm CB}+5/2}^{n+1} \\ \vdots \\ \Pi_{i_{\max}}^{n+1} \\ \Phi_{i_{\max}+1/2}^{n+1}
\end{array}\right) = {\rm RHS}. \tag{C13}$$

The last two rows of the matrix equation represent the boundary conditions (4.6) and (4.7), and the rows corresponding to $i = i_{\rm CB}+1$ and $i = i_{\rm CB}+1/2$ represent the two interpolation equations. All other rows represent Eq. (C8) or (C9) for some particular value of $i$. All quantities located on time slice $t_n$ are absorbed into the right-hand side (RHS) of the matrix equation. Nonzero entries in the square matrix are indicated by either dots or crosses. The single cross on the diagonal element of each row denotes the grid point being determined by that equation. The square matrix is band diagonal, so Eq. (C13) is easily solved by standard methods[21].

The above algorithm has the important property that grid points inside of the causal boundary cannot influence grid points in the exterior. This is true not only in the continuum limit, but in the discrete case as well. For example, since $\Pi_1^{n+1}$ appears in only one equation, it must be determined by that equation; hence it cannot possibly affect $\Phi$ or $\Pi$ at any other grid point.

Until now we have not specified the spatial averaging parameter $\theta_i$ that appears in the time derivative operator (C1). Normally one would set $\theta_i = 0$, since averaging time derivatives over space makes the difference scheme dispersive. However, this choice is inadequate for the grid points $i = 1, \ldots, i_{\rm CB}$, where we solve for $\Pi_i^{n+1}$ and $\Phi_{i+1/2}^{n+1}$ using equations centered at $i+1$ and $i+3/2$. This is because the matrix elements multiplying $\Pi_i^{n+1}$ and $\Phi_{i+1/2}^{n+1}$ in rows $i$ and $i+1/2$ are small in magnitude compared to other elements in the same row, so that when one inverts the matrix to solve for these variables, one effectively sums several terms that nearly cancel and then divides by a small number. As a result, the matrix inversion is unstable. To cure this, we set $\theta_i = 1/(2+2\lambda_i)$ inside the causal boundary $r = r_{\rm CB}$, so that there is a stronger coupling between neighboring spatial grid points. Outside the causal boundary, we do not need this coupling, so we set $\theta_i = 0$. Using different values of $\theta_i$ in the exterior than in the interior causes no problem because the two regions of the grid are causally disconnected, even in the finite difference approximation.

Because we use an implicit difference scheme, there is no stability limitation on the time step. However, for accuracy it is useful to set the time step so that inside the causal boundary, the past directed light cone of a grid point at $(i, t_{n+1})$ contains the event $(i+1, t_{n+1/2})$, as in Figure 13B. In general, this requires the time step to be larger than the grid spacing. Notice that a Courant type condition would produce entirely the opposite effect: if one decreased the time step in Figure 13B sufficiently, the past light cone of the point $(i = 3, t = t_{n+1})$ would not contain the spacetime event $(i = 4, t = t_{n+1/2})$. From Eq. (2.68) we take the center of the light cone to be at $\Delta r = -\beta\Delta t$, so we impose the condition

$$\Delta t = \varepsilon \frac{\Delta r}{\beta} \tag{C14}$$

at the innermost grid point. We typically choose $\varepsilon = 1/2$.

## APPENDIX D. AHBC SOLUTION OF CONSTRAINTS

To solve for $\psi$ and $Z$, we first find the value of $\psi$ at the AH using the evolution equation (4.12). This equation also provides an initial guess for $\psi$ elsewhere. We use the finite difference approximation

$$\mathcal{T}[\psi]_i^n = \frac{\beta_i^n}{r_i}\left(\mathcal{D}[\psi]_i^n + \frac{1}{2}\psi_i^n\right) - \frac{1}{4}\alpha_i^n \psi_i^n (K_{\rm T})_i^n, \tag{D1}$$

where the operator $\mathcal{D}$ is defined in Eq. (C6) and the operator $\mathcal{T}$ is given by Eq. (A1). This scheme is second-order accurate in space and time, even for a nonuniform grid or for unequal timesteps. The stability of the scheme is irrelevant since the result is only retained at the AH—at all other grid points, $\psi$ is refined using the Hamiltonian constraint.

Next, we solve the momentum and Hamiltonian constraints simultaneously using an iterative scheme. These equations are coupled because Brans-Dicke scalar radiation terms containing $\psi$ appear in the momentum constraint (4.8). Let $\hat{\psi}$ be an initial guess for $\psi$, and let $\hat{Z}$ be an initial guess for $Z$. If we substitute

$$\psi = \hat{\psi}\left(1 + \frac{\psi - \hat{\psi}}{\hat{\psi}}\right) \tag{D2a}$$

$$Z = \hat{Z}\left(1 + \frac{Z - \hat{Z}}{\hat{Z}}\right) \tag{D2b}$$

into the constraint equations (4.8) and (4.9), and expand to first order in the small quantities $(\psi - \hat{\psi})/\hat{\psi}$ and $(Z - \hat{Z})/\hat{Z}$, the result is

$$Z_{,\eta} + \psi\left(6\hat{\psi}^5 r^3\right)\left(\Pi_{,\eta} + \frac{\Pi\Phi\omega r}{\phi}\right) = 8\pi r^4 \tilde{S}_r + \left(5\hat{\psi}^6 r^3\right)\left(\Pi_{,\eta} + \frac{\Pi\Phi\omega r}{\phi}\right), \tag{D3}$$

$$\begin{aligned}
&\psi_{,\eta\eta} + \psi_{,\eta}\left(1 + \frac{\Phi r}{2\phi}\right) + Z\left[\frac{3}{8}\frac{\hat{Z}}{\phi^2\hat{\psi}^7 r^4}\right] \\
&\quad + \psi\left[-\frac{2\pi\tilde{\rho} r^2}{\phi\hat{\psi}^2} - \frac{21}{16}\frac{\hat{Z}^2}{\phi^2\hat{\psi}^8 r^4} + \frac{r}{4\phi}\left(\Phi_{,\eta} + 2\Phi\right) + \frac{\omega r^2}{8\phi^2}\left(\Phi^2 + 5\Pi^2\hat{\psi}^4\right)\right] \\
&\quad = -\frac{9}{8}\frac{\hat{Z}^2}{\phi^2\hat{\psi}^7 r^4} - \frac{4\pi\tilde{\rho} r^2}{\phi\hat{\psi}} + \frac{\omega\Pi^2 r^2\hat{\psi}^5}{2\psi^2}.
\end{aligned} \tag{D4}$$

Applying the same linearization procedure to the boundary condition at the AH, Eq. (4.10), we obtain

$$\psi_{,\eta} + \psi\left[\frac{1}{2} + \frac{3}{4}\frac{\hat{Z}}{\hat{\psi}^4 r^2\phi}\right] - \frac{1}{4}\frac{Z}{\hat{\psi}^3 r^2\phi} = \frac{3}{4}\frac{\hat{Z}}{\hat{\psi}^3 r^2\phi} \qquad \text{at AH.} \tag{D5}$$

For the two other boundary conditions, we set the $\psi$ at the AH to the value obtained from the evolution equation, and we use Eq. (4.11) at the outer grid point.

The constraints, together with the boundary conditions that they must satisfy, can be written in the finite difference form

$$\mathcal{D}_{1/2}[Z]_{i+1/2}+\mathcal{A}_\eta[\psi]_{i+1/2}\mathcal{A}_\eta\left[6\hat{\psi}^5r^3\right]_{i+1/2}\left(\mathcal{D}_{1/2}[\Pi]_{i+1/2}+\mathcal{A}_\eta\left[\frac{\Pi\Phi\omega r}{\phi}\right]_{i+1/2}\right)$$
$$=8\pi\mathcal{A}_\eta\left[r^4\tilde{S}_r\right]_{i+1/2}+\mathcal{A}_\eta\left[5\hat{\psi}^6r^3\right]_{i+1/2}\left(\mathcal{D}_{1/2}[\Pi]_{i+1/2}+\mathcal{A}_\eta\left[\frac{\Pi\Phi\omega r}{\phi}\right]_{i+1/2}\right), \qquad \text{(D6)}$$

$$\mathcal{D}^2[\psi]_i+\mathcal{D}[\psi]_i\left(1+\frac{\mathcal{A}_\eta[\Phi]_ir_i}{2\phi_i}\right)+Z_i\left[\frac{3}{8}\frac{\hat{Z}_i}{\phi_i^2\hat{\psi}_i^7r_i^4}\right]+\psi_i\left[-\frac{2\pi\tilde{\rho}_ir_i^2}{\phi_i\hat{\psi}_i^2}\right.$$
$$\left.-\frac{21}{16}\frac{\hat{Z}_i^2}{\phi_i^2\hat{\psi}_i^8r_i^4}+\frac{r_i}{4\phi_i}\left(\mathcal{D}_{1/2}[\Phi]_i+2\mathcal{A}_\eta[\Phi]_i\right)+\frac{\omega r_i^2}{8\phi_i^2}\left(\mathcal{A}_\eta[\Phi]_i^2+5\Pi_i^2\hat{\psi}_i^4\right)\right]$$
$$=-\frac{9}{8}\frac{\hat{Z}_i^2}{\phi_i^2\hat{\psi}_i^7r_i^4}-\frac{4\pi\tilde{\rho}_ir_i^2}{\phi_i\hat{\psi}_i}+\frac{\omega\Pi_i^2r_i^2\hat{\psi}_i^5}{2\psi_i^2}, \qquad \text{(D7)}$$

$$\frac{1}{r_{i-1/2}}\mathcal{D}_{1/2}[\psi r]_{i-1/2}=1-\frac{1}{8}\left(\Pi_ir_i+\Pi_{i-1}r_{i-1}\right), \qquad i=i_{\max}, \qquad \text{(D8)}$$

$$\mathcal{D}[\psi]_i+\psi_i\left[\frac{1}{2}+\frac{3}{4}\frac{\hat{Z}_i}{\hat{\psi}_i^4r_i^2\phi_i}\right]-\frac{1}{4}\frac{Z_i}{\hat{\psi}_i^3r_i^2\phi_i}=\frac{3}{4}\frac{\hat{Z}_i}{\hat{\psi}_i^3r_i^2\phi_i}, \qquad i=i_{\rm AH}, \qquad \text{(D9)}$$

$$\psi_i=\hat{\psi}_i, \qquad i=i_{\rm AH}. \qquad \text{(D10)}$$

Here the operator $\mathcal{D}^2$ is defined by

$$\mathcal{D}^2[Y]_i\equiv[Y_{,\eta\eta}]_i=\frac{2}{\eta_{i+1}-\eta_{i-1}}\frac{Y_{i+1}-Y_i}{\eta_{i+1}-\eta_i}-\frac{2}{\eta_{i+1}-\eta_{i-1}}\frac{Y_i-Y_{i-1}}{\eta_i-\eta_{i-1}}, \qquad \text{(D11)}$$

and the operators $\mathcal{D}$, $\mathcal{D}_{1/2}$, and $\mathcal{A}_\eta$ are defined as in Eqs. (C5)–(C7). In the above equations, the averaging operator $\mathcal{A}_\eta$ takes precedence over other operations, *e. g.*,

$$\mathcal{A}_\eta[XY^2]=\mathcal{A}_\eta[X](\mathcal{A}_\eta[Y])^2. \qquad \text{(D12)}$$

Note that $\mathcal{D}[\mathcal{D}[Y]]\neq\mathcal{D}^2[Y]$, but for an equally spaced grid,

$$\mathcal{D}_{1/2}[\mathcal{D}_{1/2}[Y]]=\mathcal{D}^2[Y]. \qquad \text{(D13)}$$

Eqs. (D6)–(D10) can be encoded into a single matrix equation:

$$
\left(\begin{array}{cccccccccccc}
\times & \bullet & \bullet & \bullet & & & & & & & & \\
& \times & \bullet & \bullet & & \bullet & & & & & & \\
& & & \ddots & & & & & & & & \\
& & & \times & \bullet & \bullet & \bullet & & & & & \\
& & & & \times & \bullet & \bullet & & \bullet & & & \\
& & & & \bullet & \times & \bullet & & \bullet & & & \\
& & & & & & \times & & & & & \\
& & & & & \bullet & \bullet & \times & \bullet & & & \\
& & & & & & \bullet & \bullet & \times & & \bullet & \\
& & & & & & & & \ddots & & & \\
& & & & & & & & \bullet & \bullet & \times & \bullet \\
& & & & & & & & & & \bullet & \times
\end{array}\right)
\left(\begin{array}{c}
Z_1 \\ \psi_1 \\ \vdots \\ Z_{i_{\mathrm{AH}}-1} \\ \psi_{i_{\mathrm{AH}}-1} \\ Z_{i_{\mathrm{AH}}} \\ \psi_{i_{\mathrm{AH}}} \\ Z_{i_{\mathrm{AH}}+1} \\ \psi_{i_{\mathrm{AH}}+1} \\ \vdots \\ Z_{i_{\max}} \\ \psi_{i_{\max}}
\end{array}\right) = \mathrm{RHS}. \tag{D14}
$$

Here crosses denote elements on the diagonal, and dots represent all other nonzero entries. The RHS matrix does not depend on $\psi$ or $Z$, but does depend on the initial guesses $\hat{\psi}$ and $\hat{Z}$.

Given an initial guess, we iterate the matrix equation (D14) until convergence, using the values of $\psi$ and $Z$ at each step as initial guesses for the next step. The matrix equation is solved by a standard band-diagonal inversion technique[21].

## APPENDIX E. AHBC SOLUTION OF LAPSE AND SHIFT EQUATIONS

The lapse and shift equations, Eqs. (4.13) and (4.14), together with the boundary conditions (4.16)–(4.18), can be written in the following finite difference form:

$$
\begin{aligned}
\mathcal{D}^2[\alpha]_i + \mathcal{D}[\alpha]_i \left(\frac{1 + 2\mathcal{D}[\psi]_i}{\psi_i}\right) &= \alpha_i \left[\frac{3}{2}\left((K_{\mathrm{T}})_i r_i \psi_i^2\right)^2 + \frac{8\pi r_i^2}{\phi_i \psi_i^2}\left(\frac{\tilde{\rho}_i + \tilde{T}_i}{2 + 3/\omega}\right)\right. \\
&\quad \left. + \frac{\omega \psi_i^4 \Pi_i^2 r_i^2}{\phi_i^2} + \frac{r_i}{\phi_i}\left(\mathcal{D}_{1/2}[\Phi]_i + 2\Phi_i\left(\frac{1 + \mathcal{D}[\psi]_i}{\psi_i}\right)\right)\right], && \text{(E1)} \\
\mathcal{D}_{1/2}[X]_{i+1/2} + \frac{3}{2}\mathcal{A}_\eta[\alpha]_{i+1/2}\mathcal{A}_\eta\left[\frac{Z}{r^3\psi^6}\right]_{i+1/2} &= 0, && \text{(E2)} \\
\alpha_i - X_i r_i \psi_i^2 \frac{1/r_i^2 - (F_1)_i - (F_3)_i}{1/r_i^2 + (F_2)_i + (F_3)_i + (F_4)_i} &= 0, \qquad i = i_{\mathrm{AH}}, && \text{(E3)} \\
\frac{1}{r_{i-1/2}}\mathcal{D}_{1/2}[\alpha r]_{i-1/2} &= 1 + \frac{1}{2}\left(\Pi_i r_i + \Pi_{i-1} r_{i-1}\right), \qquad i = i_{\max}, && \text{(E4)} \\
X_i &= \frac{(K_{\mathrm{T}})_i}{2} + \frac{\Pi_i}{2}, \qquad i = i_{\max}. && \text{(E5)}
\end{aligned}
$$

Here $X_i \equiv \beta_i / r_i$, and

$$(F_1)_i \equiv \frac{8\pi}{\phi_i \psi_i^2} \left( \tilde{\rho}_i + \frac{(\tilde{S}_r)_i}{\psi_i^2} \right), \tag{E6a}$$

$$(F_2)_i \equiv \frac{8\pi}{\phi_i \psi_i^2} \left( \frac{(\tilde{S}^r{}_r)_i}{\psi_i^4} - \frac{\tilde{T}_i}{3+2\omega} + \frac{(\tilde{S}_r)_i}{\psi_i^2} \right), \tag{E6b}$$

$$(F_3)_i \equiv \frac{\omega}{2\phi_i^2} \left( \Pi_i \psi_i^2 - \Phi_i \right)^2 - \frac{\psi_i^2 \mathcal{D}[\Pi]_i}{r_i \phi_i} + \frac{\mathcal{D}_{1/2}[\Phi]_i}{r_i \phi_i} - 2\frac{\Phi_i \mathcal{D}[\psi]_i}{r_i \phi_i \psi_i}, \tag{E6c}$$

$$(F_4)_i \equiv \frac{(K_{\rm T})_i \psi_i^2}{\phi_i} \left( \psi_i^2 \Pi_i - \Phi_i \right). \tag{E6d}$$

The operators $\mathcal{A}_\eta$, $\mathcal{D}_{1/2}$, $\mathcal{D}$, and $\mathcal{D}^2$ are defined by Eqs. (C5)–(C7) and (D11). The averaging operator $\mathcal{A}_\eta$ has precedence over other operations.

The lapse and shift equations are coupled because of the boundary condition (E3). To solve them simultaneously, we write Eqs. (E1)–(E5) as a single matrix equation:

$$\left( \begin{array}{cccccccccccc}
\times & \bullet & \bullet & \bullet & & & & & & & & \\
 & \times & & \bullet & & \bullet & & & & & & \\
 & & & \ddots & & & & & & & & \\
 & & & \times & \bullet & \bullet & \bullet & & & & & \\
 & & & & \times & & \bullet & & \bullet & & & \\
 & & & & & \times & \bullet & \bullet & \bullet & & & \\
 & & & & & \bullet & \times & & & & & \\
 & & & & & & & \times & \bullet & \bullet & \bullet & \\
 & & & & & & \bullet & & \times & & \bullet & \\
 & & & & & & & & & \ddots & & \\
 & & & & & & & & & & \times & \\
 & & & & & & & & & & \bullet & \times
\end{array} \right)
\left( \begin{array}{c}
X_1 \\ \alpha_1 \\ \vdots \\ X_{i_{\rm AH}-1} \\ \alpha_{i_{\rm AH}-1} \\ X_{i_{\rm AH}} \\ \alpha_{i_{\rm AH}} \\ X_{i_{\rm AH}+1} \\ \alpha_{i_{\rm AH}+1} \\ \vdots \\ X_{i_{\rm max}} \\ \alpha_{i_{\rm max}}
\end{array} \right) = \text{RHS}. \tag{E7}$$

Here crosses denote elements on the diagonal, and dots represent all other nonzero elements. This equation is solved for $\alpha$ and $X$ by a standard band-diagonal inversion method[21].